\documentclass[twocolumn,aps,showpacs,prc]{revtex4}
\usepackage{graphicx}
\usepackage{longtable}
\usepackage{dcolumn}
\usepackage{bm}
\usepackage{multirow}
\usepackage{color}

\begin{document}
\title{Large amplitude pairing fluctuations in atomic nuclei}

\author{Nuria L\'opez Vaquero}
\affiliation{Departamento de F\'isica Te\'orica, Universidad
  Aut\'onoma de Madrid, E-28049 Madrid, Spain}
\author{J. Luis Egido}
\email{j.luis.egido@uam.es}
\affiliation{Departamento de F\'isica Te\'orica, Universidad
  Aut\'onoma de Madrid, E-28049 Madrid, Spain}
\author{Tom\'as R. Rodr\'iguez}
\affiliation{Institut f\"ur Kernphysik, Technische Universit\"at Darmstadt, Schlossgartenstr. 2, D-64289 Darmstadt, Germany}


\date{\today}

\begin{abstract}

   Pairing fluctuations are self-consistently  incorporated on the same footing as the quadrupole deformations in present state
   of the art calculations including particle number and angular momentum conservation as well as  configuration mixing. 
The approach is complemented by the use of the finite range density dependent  Gogny force which,
with a unique source for the particle-hole and particle-particle interactions,  guarantees a self-consistent interplay in both channels.

We have applied our formalism to study the role of the pairing degree of freedom in the description of the most 
relevant observables like spectra, transition probabilities, separation energies, etc.  
We find that the inclusion 
of pairing fluctuations  mostly affects the description of excited states, depending on the excitation energy and the angular momentum. $E0$ transition probabilities experiment rather big changes while $E2$'s are less affected.  
  Genuine pairing vibrations are thoroughly  studied with the conclusion that deformations strongly
  inhibits their existence.
  
  These studies have been performed for a selection of nuclei: spherical, deformed and with different degrees
  of collectivity.

\end{abstract}

\maketitle

\section{Introduction}

The self-consistent mean field approach (MFA) with effective phenomenological interactions have succeeded in describing many bulk properties along the whole nuclear chart. This success is closely related to the incorporation of the spontaneous symmetry breaking mechanism in the mean field approach  allowing thereby  the inclusion of many correlations within a simple intrinsic product wave function (w.f.)\cite{RevModPhys.75.121}. 
 
Non-bulk observables  require the use of beyond mean field theories (BMFT).  The description of nuclear spectra and transitions, for example, requires at least one more degree  of sophistication, namely good quantum numbers. Therefore the first step to improve the MFA is the restoration  of the broken symmetries.  This is usually done by projection techniques and it is obvious that we have to recover at least the main quantum numbers, i.e., the number of particles, the angular momentum and the parity.  
This procedure is known as symmetry conserving mean field approximation (SCMFA)\cite{Mang1975325,Egido1982189,Schmid2004565}. The self-consistent implementation of the projection
is done in the variation after the projection  (VAP) approach\cite{PhysRev.135.B22,Egido198219,Anguiano2001a}, a poorer implementation is obtained in the projection after the variation (PAV) 
method. It is obvious that the former approach is much better since one varies wave functions with the right quantum numbers, the associated numerical difficulty is, however, much larger. In the case of the number of particles both approaches are in use but for the angular
momentum and with effective forces only the PAV one is tractable. 

In spite of the numerical complication the wave functions of the SCMFA behave in many  aspects as mean field ones since  fluctuations around the most probable values are ignored. 
The inclusion of this fluctuations leads us directly to the next improvement, namely, the implementation of the configuration mixing technique. This is the so-called Symmetry Conserving Configuration Mixing (SCCM)
\cite{PhysRevC.78.024309,PhysRevC.81.044311,Rodriguez-Guzman2002a,PhysRevC.81.064323} approach.This is performed within the Generator Coordinate Method (GCM) which allows to deal with collective degrees of freedom in a simple way. 
In this approach the GCM w.f. is expressed as a linear combination of basis states with different values of  chosen collective coordinates.   For the long range part of the interaction one uses as collective coordinates  the multipole expansion of the deformation operators,  namely, quadrupole, octupole, hexadecupole, etc.
For the short range part  the energy gap parameter seems to be the most appropriated collective coordinate.  Since each degree of freedom increases considerably the CPU time one  can consider explicitly only the most relevant ones.
 In practice  only the quadrupole deformation is typically considered explicitly in axial or triaxial calculations.
  However it is also well known that pairing correlations play a relevant role in the description of nuclear observables\cite{50yearsBCS}.
As a matter of fact a proper description at the mean field level is only achieved in the Hartree-Fock-Bogoliubov (HFB) approach \cite{PhysRev.122.992} which deals quadrupole and monopole degrees of freedom on the same footing. It seems therefore desirable to consider also  the deviation of the mean field values for both  degrees of freedom in a BMFT. 

Pairing fluctuations are an important degree of freedom in many nuclear phenomena and have been widely investigated \cite{Siegal197281,RevModPhys.61.131,Bes1988237,PhysRevC.36.1144,Staszczak1989589,PhysRevLett.107.092501,PhysRevC.85.044315}.

    In the gauge space associated with pairing the HFB w.f.  has two collective degrees of freedom, the pairing
gap $\Delta$, which measures the amount of  pairing correlations, i.e., the ``deformation'' \cite{NewEntry6} in the associated gauge space, and  the angle $\varphi$ which indicates  the orientation of the HFB state in this space. The HFB minimization determines  the w.f. and thereby $\Delta$   while the gauge angle $\varphi$ does not play any role at the mean field level.  The degree of freedom associated  to $\varphi$  has been exploited in the past \cite{PhysRev.135.B22}:  linear combinations of w.f.'s with different orientation in the gauge space  provide a number conserving wave function.  Pairing vibrations -associated with w.f.'s with different pairing gaps- around the average gap parameter  of the energy minimum, on the other hand, have attracted little attention. As a matter of fact they have been considered only either with very schematic interactions in the framework of the collective Hamiltonian 
\cite{Broglia_NPA_143,Gd198550,Prochniak1999181,Zajac19971}, in  microscopic model calculations \cite{PhysRevB.68.184505,mafeg05} or in reduced configuration space\cite{Faessler1973420}.  In the  approach closest to
 ours,  by Meyer et al  \cite{Meyer1991307},   pairing vibrations were considered  to study the stability of superdeformed shapes in $^{194}$Hg in  the framework of the Generator Coordinate Method  with the Skyrme  force. In this calculations, however,
neither particle number nor angular momentum projection were considered. These projections are very important 
not only to have  good quantum numbers but also because the Hamiltonian and norms overlaps are quite different
changing completely the dynamic of the system. 

 Recently, we have published a Letter  \cite{Vaquero2011520},  which simultaneously considers  the quadrupole and
 the pairing degrees of freedom in the SCCM framework with simultaneous particle number and angular momentum projections. This first study was limited  to analyze the impact of the pairing fluctuations on the spectrum of  the nucleus $^{54}$Cr.
The purpose of this paper is to extend the analysis  to other observables like  transition probabilities, separation energies, pairing vibrations, etc,  as well as to explore different parts of the chart of nuclides.

The paper is organized as follows.  In Sect.~\ref{theory} the basic theoretical formulation is provided.  In Sect.~\ref{details} the practical implementation of the theory is shown. In part \ref{pot_ener} the potential energy surfaces are discussed for the one and two dimensional case.  The norm overlaps are considered in Sect.~\ref{normas} and the energy spectra for several nuclei are displayed in Sect.~\ref{espectros}. The collective wave functions  are discussed in Sect.~\ref{coll_w_f}. The existence of genuine pairing vibration is tackled in Sect.~\ref{pair_vib} and the particles
number distribution of non particle number projected wave functions is discussed in Sect.~\ref{part_dist}. Finally
 some miscellaneous calculations are discussed in Sect.~\ref{misc}.

\section{Theoretical framework}
\label{theory}
The starting point of any BMFT is the generation of the HFB basis states in which  the GCM wave function can be expanded. According to the different coordinates  to be considered in the calculations, the HFB w.f.'s  $|\phi\rangle$
have to be calculated in each point of the grid subtended by those coordinates. This is  best achieved within  the constrained HFB approach in conjunction with the conjugate gradient method \cite{Egido199570}.  In the case of the multipole  coordinates the operators to be constrained are the corresponding  operators. For the
pairing coordinate and for finite range interactions  it has turned out  \cite{Vaquero2011520} that the  more convenient operator to be constrained is 
 $\Delta \hat{N}^2=(\hat{N}-\langle\phi||\hat{N}|\phi\rangle)^2$, $\hat{N}$ being the  number of particles, protons or neutrons.  For state independent pairing interactions, this operator is closely connected with the energy pairing gap $\Delta$~\cite{Vaquero2011520}.

In this work we will restrict ourselves to two coordinates, the axially symmetric quadrupole deformation, $\hat{Q}_{20}$ and the pairing correlations. The way to proceed in order to implement the corresponding fluctuations is the following: One generates a set of intrinsic HFB wave functions $|\phi\rangle$ with given quadrupole deformation $q$ and  "pairing deformation'' $\delta$ by solving the variational equation
\begin{equation}
\delta {E^{\prime}}[\phi(q,\delta)]   = 0  \label{min_E},
\end{equation}
with 
\begin{equation}
{E^{\prime}}= \frac{\langle\Phi|\hat{H}|\Phi \rangle} {\langle\Phi|\Phi \rangle}   
 - \lambda_q \langle \phi |\hat{Q}_{20} | \phi \rangle- \lambda_\delta \langle \phi|(\Delta\hat{N})^2|\phi \rangle^{1/2}, \label{E_prime}
\end{equation}
and the Lagrange multipliers $\lambda_q$ and $\lambda_\delta$ being determined by the constraints  
\begin{equation}
\langle \phi |\hat{Q}_{20} | \phi \rangle =q, \;\;\;\; \;\;\;\;      \langle \phi|(\Delta\hat{N})^2|\phi \rangle^{1/2} = \delta,
\end{equation}
 It remains to specify the meaning of the w.f. $|\Phi\rangle$. One has to distinguish several cases: In the more elaborated approach 
 $|\Phi\rangle=\hat{P}^{N}...\hat{P}^{I} |\phi\rangle$ with $\hat{P}$ 
 an operator which restores the symmetry (symmetries) broken by the intrinsic w.f. $|\phi\rangle$. In this case  we are solving the self-consistent SCMFA, i.e., in the VAP approach. In our  case we are interested in restoring the rotational invariance and the number of particles symmetry, i.e., we should take
  $|\Phi \rangle=  \hat{P}^N \hat{P}^I |\phi\rangle$,  where $\hat{P}^N$  performs the particle number projection (PNP) and 
$\hat{P}^I$ the angular momentum projection (AMP). This is the most general SCMFA and it is very CPU time consuming. 
However, taking into account that, in general, the semiclassical approximation of using Lagrange parameters is a very good
approximation for strong collective symmetry breaking as in the case of the angular momentum -and less 
 appropriated for non very collective symmetry like the particle number-  we shall
 consider   $|\Phi\rangle=\hat{P}^N|\phi\rangle$. That means, in this case only the particle number restoration is considered in the VAP approach whereas the angular
 momentum projection is performed after the variation has taken place, i.e., in the PAV approach. We shall call this procedure PN-VAP
 approach. This method although  more involved provides a much better description of the pairing correlations in the intrinsic w.f. \cite{Anguiano2001a}.
 In the simplest case
  $|\Phi\rangle\equiv|\phi\rangle$ and, therefore,   we are solving the plain HFB equations. Notice that in this case we have to add, to the right hand side of Eq.~\ref{E_prime},  the term $-\lambda_{N} \hat{N}$ in order to keep the right number of particles   on the average,  $\lambda_{N}$  is fixed by the constraint $\langle\phi | \hat{N}| \phi\rangle=N $. In this case both the PNP and the AMP are performed in the PAV approach.
  
Once we have generated the basis states we can proceed with the configuration mixing calculation. This is carried out within the generator coordinate method taking linear combinations of the wave functions obtained in the first step after performing projections on the required symmetries
\begin{eqnarray}
|\Psi^{b,\sigma}\rangle=\int f^{b,\sigma}(q,\delta) \; \hat{P}^{\rm B}\;|\phi(q,\delta)\rangle dq d\delta.
\label{GCMstate}
\end{eqnarray} 
Where $b$ denotes the set of quantum numbers to which $\hat{P}^{\rm B}$ projects on, in
the most general case $b=I,N,Z$ \footnote{In order to simplify the notations in this paper we will not distinguish between protons and neutrons thus, for example, when we write  $\hat{P}^B=  \hat{P}^N\hat{P}^I$ we mean   $\hat{P}^{\rm B}=  \hat{P}^Z\hat{P}^N\hat{P}^I$, with $Z$ and $N$ the
 proton and neutron numbers respectively.}.  
Then, the variational principle applied to the weights $ f^{b,\sigma}(q,\delta)$ gives the generalized eigenvalue problem, the Hill-Wheeler (HW) equation:
\begin{eqnarray}
\!\!\!\!\!\!\!\!\!\!\!\!\!\!\!\int \left(\mathcal{H}^{b}(q \delta,q^\prime \delta^\prime)-E^{b,\sigma}\mathcal{N}^{b}( q \delta,q^\prime \delta^\prime)\right)f^{b,\sigma}(q^\prime \delta^\prime)dq^\prime d\delta^\prime=0,
\label{HWeq}
\end{eqnarray}
with  $\mathcal{H}^{b}$ and $\mathcal{N}^{b}$ the  Hamiltonian and norm  overlaps, respectively and $\sigma=1,2, ...$, numbers the ground and excited states with quantum numbers  $b$, $E^{b,\sigma}$ is the energy of the corresponding state,  $\sigma=1$  being the Yrast state. 
  See Ref.~\cite{Rodriguez-Guzman2002a,PhysRevC.81.064323} for a detailed description of the solution of the HW equation. The collective wave functions are provided by \cite{ring}
 \begin{equation}
g^{b,\sigma}(q\delta)  =  \sum_{q^{\prime}\delta^{\prime}}{{\cal N}^b}^{1/2}(q\delta,q^{\prime}\delta^{\prime}) f^{b,\sigma}(q^{\prime}\delta^{\prime}), 
\label{Coll_wf} 
\end{equation}
which are orthogonal and  satisfy $\sum_{q\delta} |g(q\delta)^{b,\sigma}|^2 = 1$, the weights $f^{b,\sigma}(q\delta)$, on the other hand, do not satisfy these properties.

In the  case that  in Eq.~\ref{GCMstate}  the operator $\hat{P}^{\rm B}= \hat{P}^I$, i.e., without PNP, since the constraint on the particle number is done at the HFB level,  nothing guarantees that   $\hat{P}^I|\phi(q,\delta)\rangle$ nor $|\Psi^{I,\sigma}\rangle$ do have the right values for the number of protons and neutrons. In order to correct for this deficiency the usual cranking recipe \cite{ring} of minimizing $ \hat{H}^{\prime}=H -\lambda \hat{N}$ instead of ${H}$ is used. This amounts to substitute $ \mathcal{H}$  by $ \mathcal{H}^{\prime}$ in Eq.~\ref{HWeq} , with $ \mathcal{H}^{\prime}$ given by
\begin{equation}
(\mathcal{H}^{\prime})^{I}(q \delta,q^\prime \delta^\prime)
= \mathcal{H}^{I}(q \delta,q^\prime \delta^\prime)-\lambda {\Delta N}^{I}( q \delta,q^\prime \delta^\prime)
\label{lamb_correct}
\end{equation} 
the parameter $\lambda$ is an average of the Lagrange parameters $\lambda_N$ of the HFB equations in the different $(q \delta)$ points and ${\Delta N}^{I}( q \delta,q^\prime \delta^\prime)=\langle \phi( q \delta)|(\hat{N}-N)P^I |\phi(q^\prime \delta^\prime)\rangle$ \cite{Bonche1990466}.

\section{Details of the calculations}
\label{details}

We use in our numerical application the finite range density dependent Gogny force with the D1S parametrization \cite{Berger198423}. A detailed description of the evaluation of the different terms of the Hamiltonian in the HFB and in the PN-VAP can be found in \cite{Anguiano2001a} and in the AMP and GCM in \cite{Rodriguez-Guzman2002a,PhysRevC.81.064323}. The density dependent term of the interaction has been scrutinized in the last decade \cite{Anguiano2001a,PhysRevC.79.044319,PhysRevC.76.054315,PhysRevC.84.014309}, we  therefore provide a short discussion
of our approach to this term in the Appendix. In order to have a tractable problem we limit our  configuration space  to eight spherical\cite{Egido1993253} oscillator shells.  Though we will concentrate on small nuclei, this configuration space, in some cases,  might be too
small to provide quantitative results.  As already mentioned we limit ourselves also to axially symmetric deformations.

We present three different methods based on different types of projections explained above, namely,  HFB+AMP, HFB+PNAMP and PN-VAP+PNAMP.
In the first one (HFB+AMP) particle number projection is completely ignored, i.e., the intrinsic w.f., $|\phi(q,\delta) \rangle$, is determined in the HFB approach and afterwards AMP is performed in the configuration mixing procedure, in this case $P^B=P^I$. In the second one, (HFB+PNAMP) $|\phi(q,\delta) \rangle$ is determined in the HFB approach but then PNP and AMP are performed in the configuration mixing method, i.e, 
 now $P^B=P^I P^N$. In the last one (PN-VAP +PNAMP),  $|\phi(q,\delta) \rangle$ is determined in the PN-VAP method and afterwards the PNP and AMP are performed in the configuration mixing mechanism.
Furthermore in order to investigate the effect of the pairing fluctuations we shall present results in one dimension (1D), i.e., only with the quadrupole moment as coordinate generator and in two dimensions (2D) with the pairing energy and the quadrupole moment as generator coordinates.  

In principle the calculations should be 3D with coordinates $(q, \delta_Z, \delta_N)$ with separate constraints for neutrons and protons:
\begin{equation}
\langle \phi|(\Delta\hat{N})^2|\phi \rangle^{1/2} = \delta_N, \;\;\;\; \;\;\;\;      \langle \phi|(\Delta\hat{Z})^2|\phi \rangle^{1/2} = \delta_Z,
\end{equation}
unfortunately with three constraints  the problem becomes computationally very demanding.  What we have done is to substitute
the above constraints by a single one on $\delta$, the Lagrange multiplier $\delta$ being defined by:
\begin{equation}
\langle \phi|(\Delta\hat{N})^2|\phi \rangle^{1/2}  +   \langle \phi|(\Delta\hat{Z})^2|\phi \rangle^{1/2} = \delta,
\end{equation}
In order to check the quality of our approximation we have performed three kind of  calculations, in the first one we have constrained only $\delta_N =\langle \phi|(\Delta\hat{N})^2|\phi \rangle^{1/2}$, the variational principle guarantees  that in this case 
$\langle \phi|(\Delta\hat{Z})^2|\phi \rangle^{1/2}$ self-consistently will take the optimal value as to minimize the energy. In the
second one we constraint only on $\delta_Z = \langle \phi|(\Delta\hat{Z})^2|\phi \rangle^{1/2}$ whereas $\langle \phi|(\Delta\hat{N})^2|\phi \rangle^{1/2}$ is allowed to vary freely and in the third one we constraint on 
$\delta =\langle \phi|(\Delta\hat{N})^2|\phi \rangle^{1/2}+ \langle \phi|(\Delta\hat{Z})^2|\phi \rangle^{1/2}$. We have performed 
calculations in the PN-VAP+PNAMP approach for the nucleus ${^{50}}$Ca, we have chosen this nucleus because it is magic in
protons and in this case it would be clearer to decide about the quality of the approaches.  The ground state energies are -428.991 MeV (constraint on $\delta_Z$), -428.962 MeV  (constraint on $\delta_N$) and -429.037 MeV  (constraint on $\delta$). 
Though the lowest energy is obtained in the case where we  constraint on $\delta$ we do not find significant differences in the three calculations. Concerning the excitation energies and other relevant observables we present in Table~\ref{table1} the excitation 
energies of the three low-lying states for angular momentum $0^+$ and $2^+$ and the proton and neutron pairing energies in the three mentioned calculations, see \cite{Anguiano2001a} for the definition of the pairing energy. 
A close look to these results shows again that there are not remarkable differences.
Besides, the same type of calculations for the nucleus $^{32}$Mg reconfirm these findings.
 In conclusion the constraint in $\delta$ is appropriate as one does not look for specific  proton or neutron observables.

\begingroup
\squeezetable
\begin{table}[h]
\hspace{-1cm}

\begin{tabular}{|c|c|c|c|}
\cline{1-4}
\multicolumn{1}{|c|}{} & $\delta_{Z}$ & $\delta_{N}$&  $\delta$ \\  \cline {2-4}
   &$E^*\;\;\;\;E^Z_{pair}\;\;\;\;E^N_{pair}$& $E^*\;\;\;\;E^Z_{pair}\;\;\;\;E^N_{pair}$ & $E^*\;\;\;\;E^Z_{pair}\;\;\;\;E^N_{pair}$ \\ \hline 
$0^+_{1}$& $0.00 \;-4.1 \;-5.9$ &$0.00 \; -4.2\; -6.2$ &$0.00\; -4.0\; -5.7$\\ \cline {1-4} 
$0^+_{2}$& $4.58 \;-5.0 \;-5.9$ &$4.80 \; -4.5\; -6.0 $&$4.65\; -4.7\; -5.7$\\ \cline {1-4} 
$0^+_{3}$& $7.80 \;-5.1 \;-4.7$ &$6.76 \; -4.7\; -4.3$ &$7.17\; -4.1\; -3.7$ \\ \cline {2-4}  \hline \hline
$2^+_{1}$& $2.60 \;-4.0 \;-4.0$ &$2.11 \; -4.0\; -3.1$ &$2.21\; -3.5\; -3.2$\\ \cline {1-4} 
$2^+_{2}$& $5.75 \;-4.7 \;-5.2$ &$5.31 \; -4.5\; -3.3 $&$5.59\; -4.2\; -4.2$\\ \cline {1-4} 
$2^+_{3}$& $6.05 \;-4.2 \;-4.0$ &$5.88 \; -4.6\; -4.9$ &$5.90\; -3.7\; -4.0$ \\ \cline {2-4}  \hline \hline

\end{tabular}
\caption{ Excitation energy $E^*$, proton (neutron) pairing energy $E^{Z(N)}_{pair}$ for the three lowest $0^+$ and $2^+$ states of the HW equation in the PN-VAP+PNAMP constraining on $\delta_Z$, $\delta_N$
or $\delta$.}
\label{table1}
\end{table}
\endgroup

For the $q$ coordinate we take an interval $-220$ fm$^2$ up to $+400$  fm$^2$ with  $\Delta q=20$ fm$^2$, for the pairing coordinate $\delta$ from $0$ to $4.5$ with a $0.5$ interval. The chosen $\delta$ interval covers \cite{Vaquero2011520} a pairing energy range from  $0.0$ up to $\sim 50$ MeV, to compare with typical values of a few MeV's in  the  nucleus we shall analyze. We have, therefore,  a grid with 32 points for the unidimensional calculations ($q$ coordinate) and with 320 for the two-dimensional, $(q,\delta$), ones.

\begin{figure}[h]
\begin{center}
\includegraphics[angle=0, scale=0.6]{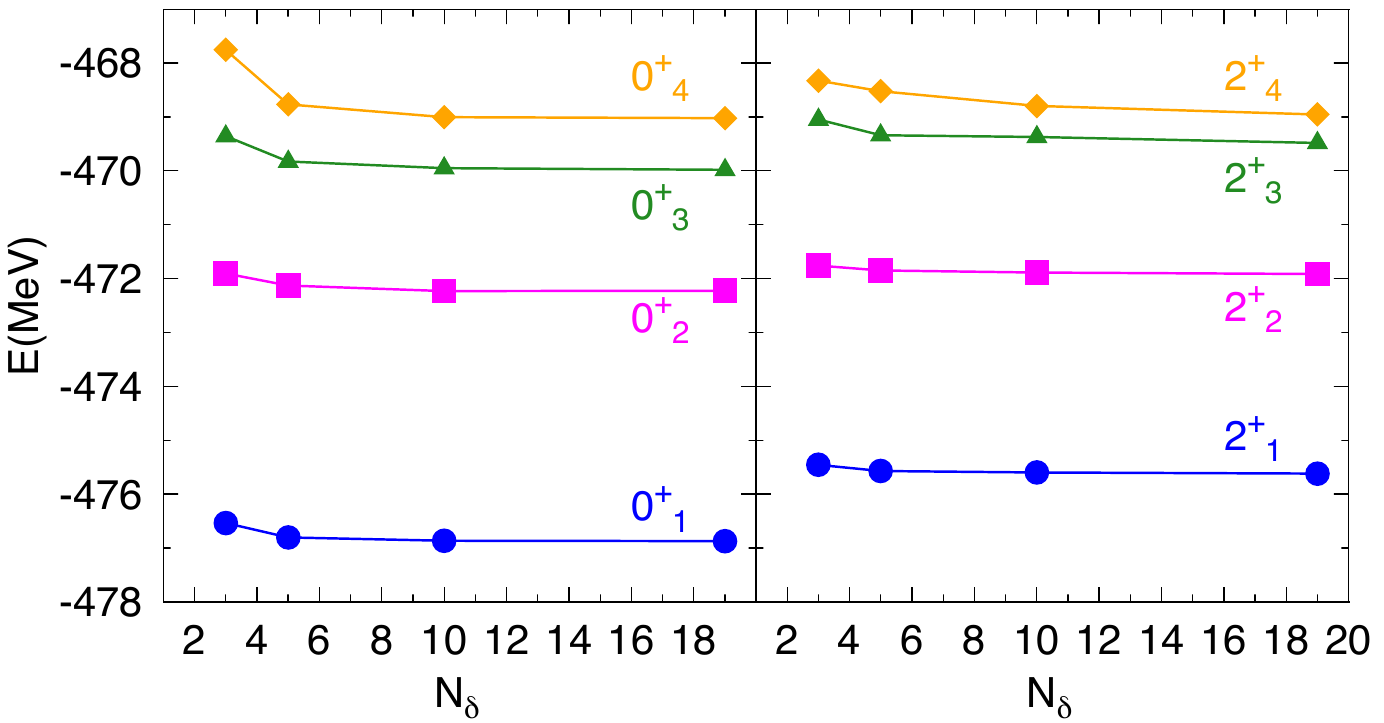}
\end{center}
\caption{(Color Online) Energies of the lowest $0^{+}$ and $2^{+}$ states as a function of the number of $\delta$ points  (3, 5, 10 or 19) used in the calculations in the PN-VAP+PNAMP approach for  $^{54}$Cr. The lines connecting the points are to guide the eye. }
\label{fig:convergence}
\end{figure}

The size of  the grid and the size interval in $q$ has been discussed in the past. As we are introducing a new coordinate,  we are interested in knowing the suitable size of the basis concerned to the $\delta$ degree of freedom. 
In Fig.~\ref{fig:convergence} we show solutions of Eq.~\ref{HWeq} calculated in the PN-VAP+PNAMP approach in 2D for the  nucleus $^{54}$Cr for different number, N$_\delta$, of delta values (3, 5, 10 or 19) in the interval $0$ to $4.5$.  We can conclude comparing these results that ten values of delta for each deformation is a good election for our calculations to guaranty a good energy convergence. Furthermore, the tails of the collective wave functions go to zero inside the interval border.    

\begin{figure}[t]
\begin{center}
\includegraphics[angle=-90, scale= 0.4]{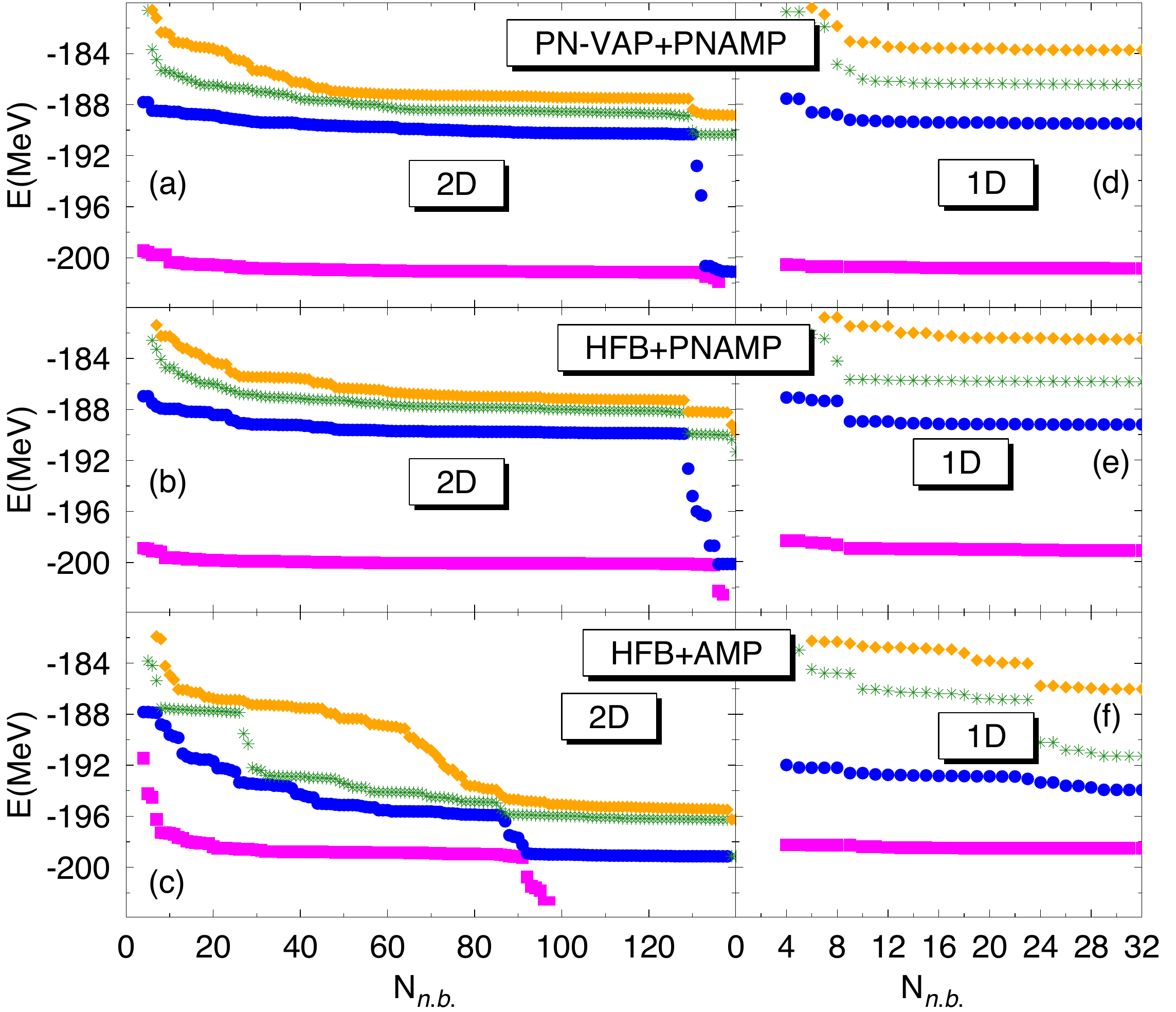}
\end{center}
\caption{(Color online) Energies of the four lowest $0^+$ states of the nucleus $^{24}$Mg. Left (right) panels correspond to the 2D (1D) calculation  both of them have been performed with the 3 different approaches. } 
\label{fig:plateau_24Mg}
\end{figure}

\begin{figure}[h]
\begin{center}
\includegraphics[angle=-90, scale=0.4]{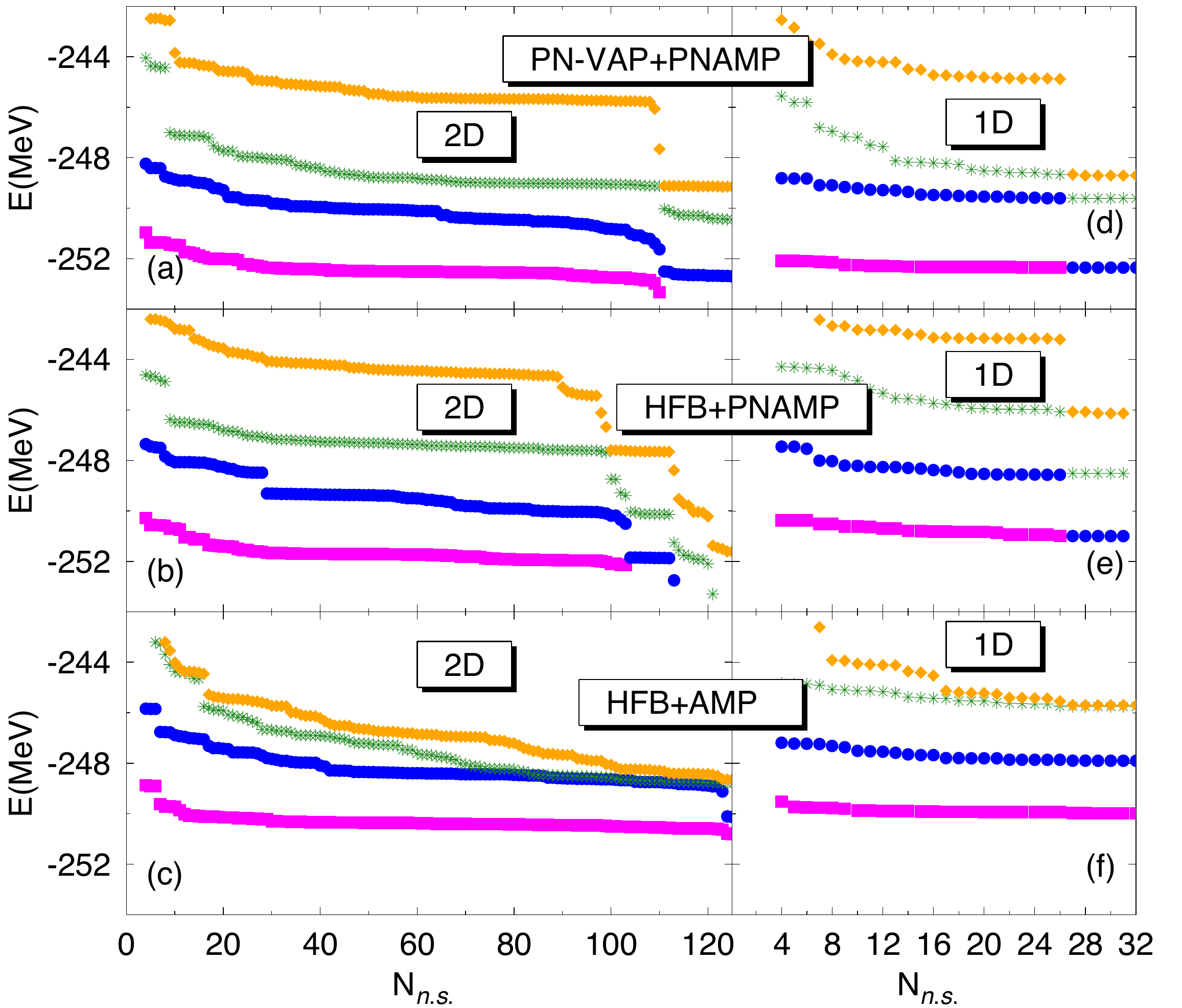}
\end{center}
\caption{(Color Online) The same plot as the above one, but for the $^{32}$Mg.  }
\label{fig:plateau_32Mg}
\end{figure}

 An inherent difficulty to the solution of the HW equation (Eq.~\ref{HWeq} is the fact that the basis states $\phi(q\delta)$ are not orthogonal.
To overcome this problem \cite{ring} one works in the natural basis where the linear dependent states do have zero or very small
norm. The natural states with a norm smaller than a given cutoff are omitted in the final diagonalization of the HW equation. The determination of the cutoff  is not always easy  and one ends up studying the convergence of the energy and the corresponding w.f. as a function of the number of the natural basis states, $N_{n.s.}$, taken in the calculations. The signal for the convergence is  the presence of  a region where the energies are nearly constant, i.e. the so called plateaus \cite{Bonche1990466}.
We now study  the convergence of the four lowest $I=0^{+}$  states  in the different approximations. 
In Figs.~\ref{fig:plateau_24Mg}  and \ref{fig:plateau_32Mg} we show the convergence for the nuclei $^{24}$Mg and  $^{32}$Mg, respectively. We represent the energy,  for the one (1D) and the two dimensional (2D) cases, versus  $N_{n.s.}$. The final choice for 
$N_{n.s.}$ is a value around which we observe a wide plateau for all the levels of interest and the collective wave functions
do not change. This value must be kept constant for a given angular momentum in order to guarantee the orthogonality of the corresponding wave functions. \\
In the actual calculations the natural states are ordered by decreasing norm eigenvalues and
the general behavior that one expects for the convergence as the number of natural basis
states increases is the following. As long as the norm eigenvalues are relatively  large  one expects
a lowering of the energy with increasing number of natural states since the Hilbert space gets larger.
We also expect faster convergence for low lying energy eigenvalues.
As the norm eigenvalues get smaller their contribution to the energy become very small and a kind of
plateau is expected.  However 
since  the linear dependent states correspond to very small norm eigenvalues one must be aware because by increasing the number of natural states one definitively reaches  a point  with such  small norms (close to zero) that the energy tends to very large negative values.    This is the general behavior and depending of the linear dependence and other aspects one can have some variations as it can be seen in the depicted cases.
   The general behavior is clearly illustrated in Fig.~\ref{fig:plateau_24Mg}(a).
For small number of natural states the lowering of the energy is considerable, the
ground state and the first excited state are converging very fast.  Taking into account  60 natural states we obtain a rather
good convergence for all four energy eigenstates.  The consideration of  additional
natural states, up to about 120, provides very small changes in the energy. Finally by $N_{n.s}\approx 130$ the norm of these states gets that small that its contribution to the energy
is spurious.  One observes that one energy eigenvalue decreases more and more as new natural states are considered. 
This spurious state crosses the physical ones (one should not get confused by the colors provided by the plotting code
which assigns a given color to the lowest energy, magenta in this case, blue to the second and so on) and becomes more
and more negative.
By $N_{n.s.}\approx 140$ this spurious state provides not only huge energy but also  huge
 values for any other observable one tries to calculate. For very large linear dependence of the basis 
 further spurious states  can be obtained for even larger $N_{n.s.}$. In the opposite situation of small basis with small 
 or none linear dependence, as in the one dimensional case, either non spurious states appear, as in 
 Figs.~\ref{fig:plateau_24Mg} d), e) and f) or the spurious state appears as  a degenerated energy eigenvalue which  never gets very large, see  Figs.~\ref{fig:plateau_32Mg} f). \\
 In the 2D case we find very wide plateaus in the PN-VAP+PNAMP and in the HFB+PNAMP cases confirming a very good convergence and not very good plateaus  for the HFB+AMP approach indicating a poorer convergence. 
These plots show that in the absence of PNP there is more mixing and larger linear dependence. We will see later on that this peculiarity appears also for other properties.  Concerning the 1D case the situation is similar to the 2D one. It appears to be less dramatic but this is due to the fact that the linear dependence is smaller  because of  the smaller size of the basis. For angular momentum larger than zero the situation is very similar to the one depicted here.  The behavior of the HFB+PNAMP approach  is not specific of the nucleus $^{24}$Mg as can be seen  in Fig. ~\ref{fig:plateau_32Mg} where we display the convergence  for the nucleus $^{32}$Mg. We find that the conclusions drawn for  $^{24}$Mg do  also apply for $^{32}$Mg. 

\section{Potencial energy surfaces}
\label{pot_ener}
In this and in the next sections we discuss several relevant properties for a variety of nuclei,  deformed, spherical, collective, etc. 
\subsection{ One dimensional case}

\begin{figure}[h]
\begin{center}
\includegraphics[angle=0, scale=0.75]{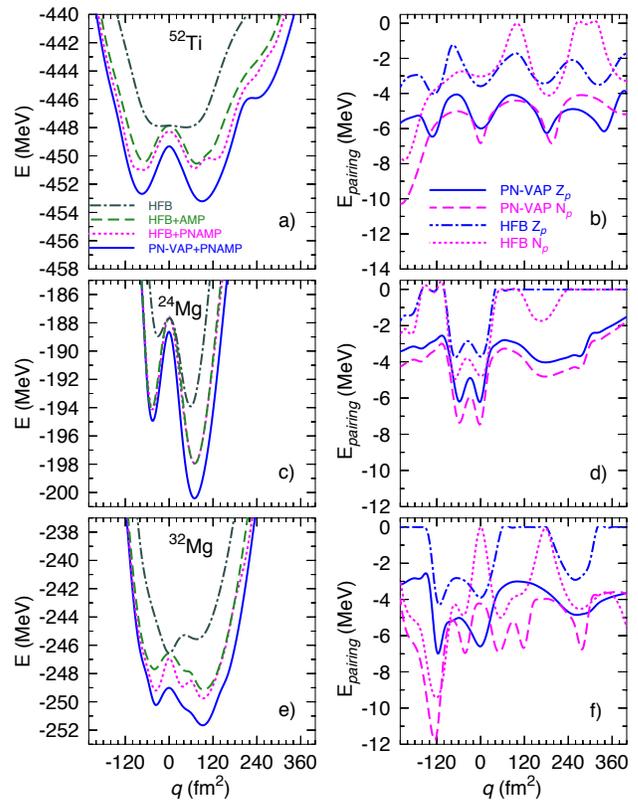}
\end{center}
\caption{(Color online) Left panels: One dimensional potential energy curves for the nuclei $^{52}$Ti,
 $^{24}$Mg and   $^{32}$Mg in the HFB and in several  approaches for $I=0^+$. Right panels: Proton and neutron  pairing energies of the intrinsic wave functions in the HFB and
 PN-VAP for the same nuclei.}
\label{fig:1D}
\end{figure}

A useful quantity for the interpretation of  observables is the diagonal of the Hamiltonian overlap entering into the HW matrix Eq.~\ref{HWeq},  as a function of the generator coordinate(s). In the 1D case we are concerned with $\langle\phi(q)|HP^B|\phi(q)\rangle/\langle\phi(q)|P^B|\phi(q)\rangle$. Where $P^B\equiv P^I$ or  $P^B\equiv P^IP^N$  depending on the approach. In Fig.~\ref{fig:1D}, left panels,  we present
the potential energy curves for the nuclei $^{52}$Ti, $^{24}$Mg and $^{32}$Mg for $I=0 \hbar$ versus the quadrupole moment.  

In the PN-VAP+PNAMP approach the  nucleus $^{52}$Ti presents coexistent prolate and oblate shapes and a prolate  super-deformed shoulder. 
  In the  $^{24}$Mg case  we obtain a strongly prolate deformed minimum and about 5 MeV above an oblate one. For $^{32}$Mg we obtain a prolate deformed minimum and an oblate pocket about 2 MeV above.
  These three nuclei correspond to  scenarios representative for the different cases one may find.
  The HFB+PNAMP and HFB+AMP curves display more or less similar shapes to the PN-VAP+PNAMP ones but lie a few MeV above. In all cases the energy gain of the PN-VAP+PNAMP approach is the largest one.  In the right panels of the same figure the
 pairing energies of the intrinsic wave functions in the HFB and in the PN-VAP approach are displayed. In the $^{52}$Ti nucleus we
 find a oscillatory behavior as a function of the deformation. In the HFB approach we observe a collapse of the neutron pairing correlations at the $q$-values corresponding to the prolate minimum and the super-deformed pocket. In the PN-VAP approach we do not observe any collapse but we obtain larger absolute values. The same trend is observed in the Magnesium isotopes: pairing collapse at the minima for the HFB approach and much larger values for the PN-VAP one. The fact that we do not obtain the same correlations for  the $N=Z$,  $^{24}$Mg nucleus has to do with the Coulomb anti-pairing effect (CouAP) \cite{Anguiano2001}. 

 \subsection{Two dimensional case}
\label{PES_2D}
\begin{figure}[h]
\begin{center}
\includegraphics[angle=0, scale=0.62]{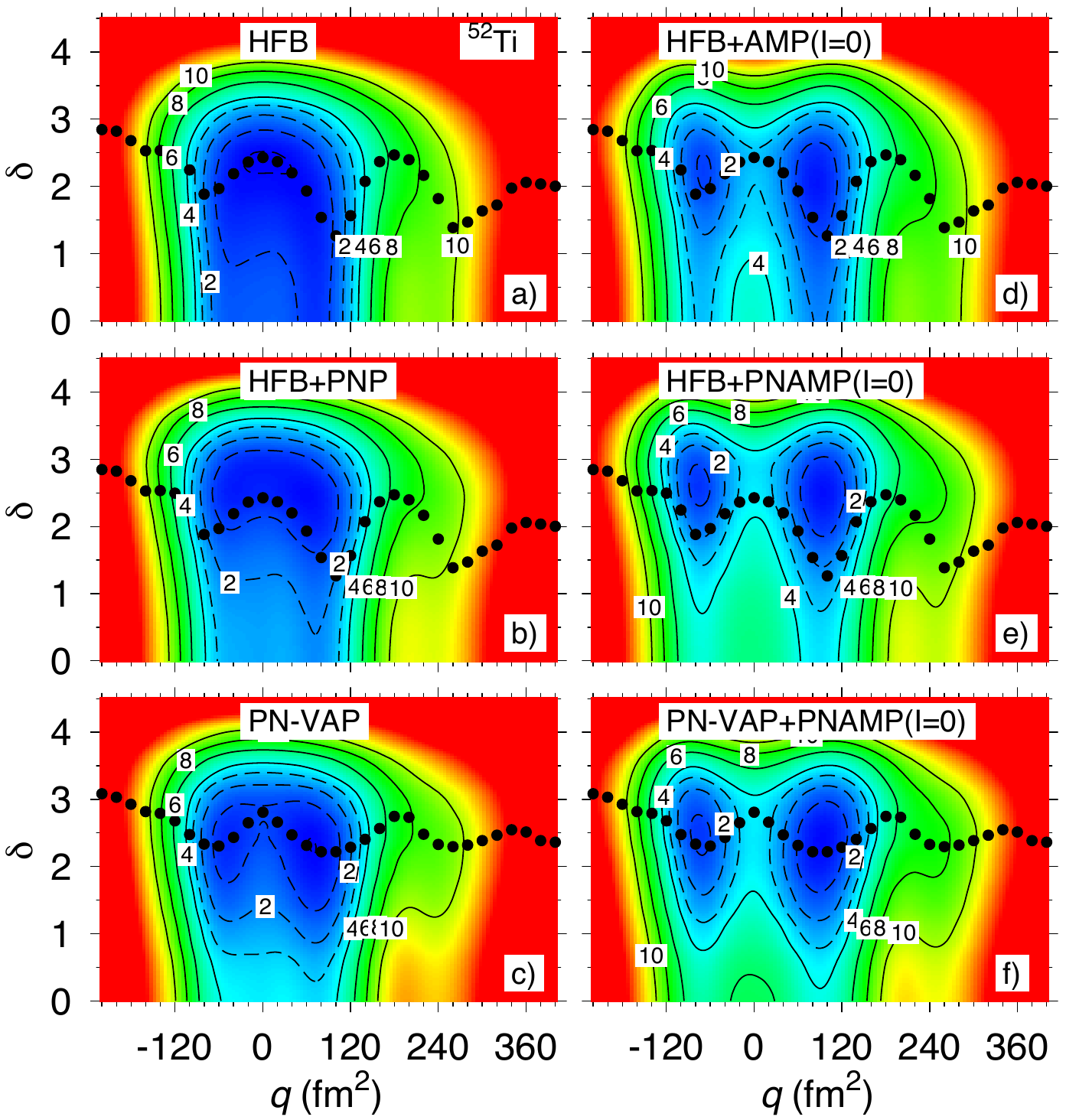}
\end{center}
\caption{(Color online) Potential energy contour plots for $^{52}$Ti in the $(\delta,q)$ plane in different approaches.  In dashed lines, equipotential lines from 0 to 3 MeV in step of 1 MeV. In continuous lines, contours from 4 to 10 MeV in steps of 2 MeV. In each panel the energy origin has been chosen independently  and the energy minimum  has been set to zero. The bullets in each panel represent the  $\delta$ values of the self-consistent solution (HFB or PN-VAP) extracted from the 1D ($q$- constrained) approach and are displayed as a discussion guide. Since all HFB based approaches do have the same intrinsic w.f. all of them have the same bullets pattern. The same apply to all PN- VAP based approaches.}
\label{fig:pairfluc_Ti52}
\end{figure}

  We study now the  dependence of the potential energy of these nuclei with respect to the two collective degrees of freedom $(q, \delta)$. In Fig.~\ref{fig:pairfluc_Ti52},  we present contour lines of the potential energy of $^{52}$Ti   as function of the constrained parameters $(q,\delta)$  in different approximations. The bullets represent the $\delta$ values of the self-consistent solutions (HFB or PN-VAP, i.e., without AMP) of the 1D ($q$-constrained) approach. They must  be orthogonal to the equipotentials curves in the corresponding approach.  The 1D plots of Fig.~\ref{fig:1D} can be used as a guide in the interpretation of the 
2D plots.

The relationship between the parameter $\delta$ and the pairing energy is rather independent of the $q$-value, see 
Fig.~2 of \cite{Vaquero2011520}. To have a feeling,  for the nucleus $^{52}$Ti and for $q=100$ fm$^2$ in the VAP+PNAMP approach and for $I=0$,  we provide the pairing energy (in parenthesis and in MeV)  corresponding to the preceding $\delta$ values: $0.0 (0.00)$, $0.5 (-0.52)$, $1.0 (-2.11)$, $1.5 (-4.74)$, $2.0 (-8.19)$, $2.5 (-12.53)$, $3.0 (-18.33)$, $3.5 (-26.17)$, $4.0 (-36.71)$, and $4.5 (-50.26)$. We thus see that the $\delta$ range covers a wide  energy interval.

In Fig.~\ref{fig:pairfluc_Ti52} (a)  we display the pure HFB case. Here  we find a region delimited
 from $q=-60$ fm$^{2}$ to $q$=100 fm$^{2}$ in the $X$ axis and from $\delta$=0 to $\delta$=2.5 in the ordinate, where the potential is soft in both directions. That means, for a given value of $q$ (or $\delta$) one does not gain much energy (just around 1 MeV) by increasing the $\delta$ coordinate (or $q$). However, for the same  $q$ interval but $\delta$ between 2.5 and 4 it takes a considerable amount of energy to increase the pairing correlations of the system. For higher values of $\delta$, the pairing energy gain is huge and the total energy is up to 20 MeV larger. An analogous conclusion is obtained for the region $-140$ fm$^{2}$ $< q< -60$ fm$^{2}$ and 120 fm$^{2}$ $< q<$ 240 fm$^{2}$,  the potential becomes stiff and to deform the nucleus to that values requires a large amount of energy.   This structure is consistent with
 the one dimensional plot shown in the top left panel of Fig.~\ref{fig:1D}.

 \begin{figure}[th]
\begin{center}
\includegraphics[angle=0, scale=0.62]{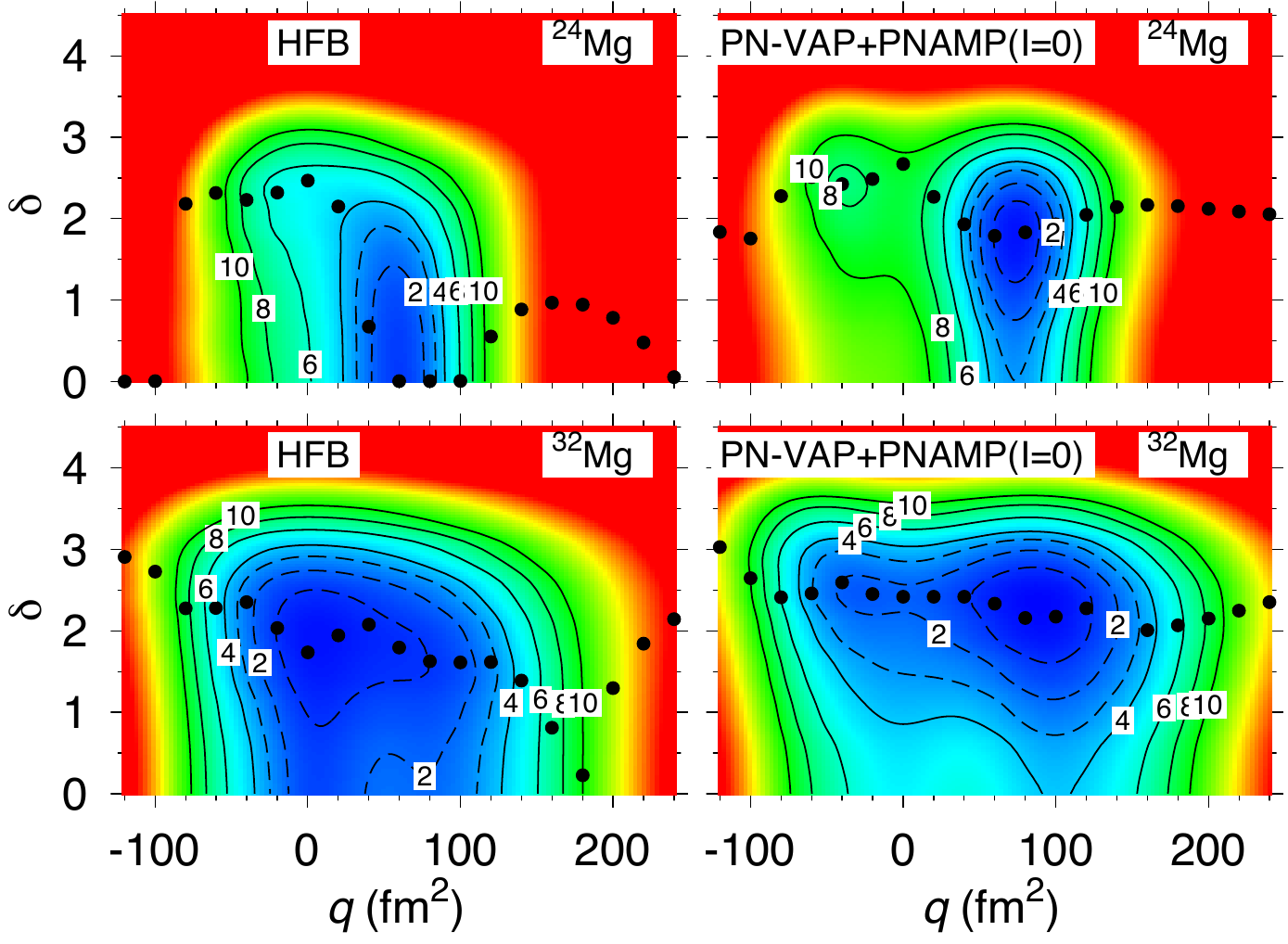}
\end{center}
\caption{Color online. Left panels: Potential energy surfaces for the nuclei  $^{24}$Mg and  $^{32}$Mg in the HFB approach. Right panels: Potential energy surfaces for the nuclei  $^{24}$Mg and  $^{32}$Mg in the PN-VAP+PNAMP  approach for I=0. In both cases the contours follow the same interval as in Fig.~\ref{fig:pairfluc_Ti52}}
\label{fig:pairfluc_Mg24_32}
\end{figure}

Next, in panel (b) we show the effect of particle number projection after the variation, i.e., one takes the HFB wave functions used to generate panel (a) and calculates the PNP energy. One obtains again rather flat minima  but  displaced to $\delta$=2.5. The energy lowering  of the absolute minimum is 1.37 MeV. 
In panel (c) we also represent the effect of the PNP but in this case, the projection is performed before the variation, therefore in that approximation we obtain the energy  with PN-VAP intrinsic wave functions. This plot looks like the previous one and the two trends mentioned  before are present here too: the equipotencials are shifted towards large values of $\delta$ and the minimum is deeper, being now even lower,  1.17 MeV below the PAV absolute minimum. One now observes two minima, one prolate at $(q=60$ fm$^{2}$, $\delta$=2.5) and one oblate at $(q=-40$ fm$^{2}$, $\delta=2.5)$. The PN-VAP approach is the proper way to perform the variation because one minimizes the energy calculated with the right number of particles.  One have to have in mind that, even though the PNP brings the energy minimum of the HFB solution closer to the VAP one,  there are other observables which values  do not coincide with the self consistent ones provided by the VAP approach.

Now, the angular momentum projection ($I=0\; \hbar$) is performed for the approaches of the left panels  and presented in the corresponding right panels.
We start with the HFB+AMP case, panel (d), here, as mentioned above, see Eq.~\ref{lamb_correct}, in each $(\delta, q)$ point we have performed the particle number energy correction due to the fact that the AMP wave function does not provide, on the average, the right particle number. As seen in Fig.~\ref{fig:1D} the AMP increases considerably the depth of the potentials and the $q$-values of the minima.  They move  to larger $q$-values, $-80$ fm $^{2}$ for the oblate minimum and $80$ fm $^{2}$ for the prolate one.   In the HFB+PNAMP, panel (e),  or in the PN-VAP+PNAMP, panel (f),  the effect of the AMP is also to widen the equipotentials and to deepen the minima, the prolate being shifted towards larger value, 100 fm$^{2}$, and the oblate one to 
 $-80$ fm$^{2}$. An interesting point is that in the 2D plot we find that the minima of the energy in the HFB+AMP approach correspond  to  pairing energies of $\delta\approx 2.0$. We find that this is not the case in the PNP cases where the minima correspond to $\delta\approx 2.5$.  The energy difference corresponding to the different $\delta$ values amounts to a difference of pairing energies of a few MeV's \cite{Vaquero2011520}. 
 The equipotential surfaces  of panels (e) and (f) look very similar though in detail they are different, c.f.  Fig.~\ref{fig:1D}.   The fact that  the minima of the HFB+AMP approach lie at a weak
 pairing region will have important consequences since the masses associated to the dynamics of the system, i.e., the solution of the HW equation, will be much larger than the ones associated to the PN projected approaches, providing a more compressed spectrum.  The energy gain of the absolute minimum in the PN-VAP+PNAMP approach with respect to the HFB (PN-VAP) is 4.53 MeV  (2.711 MeV). 

\begin{figure}[t]
\begin{center}
\includegraphics[angle=0, scale=0.62]{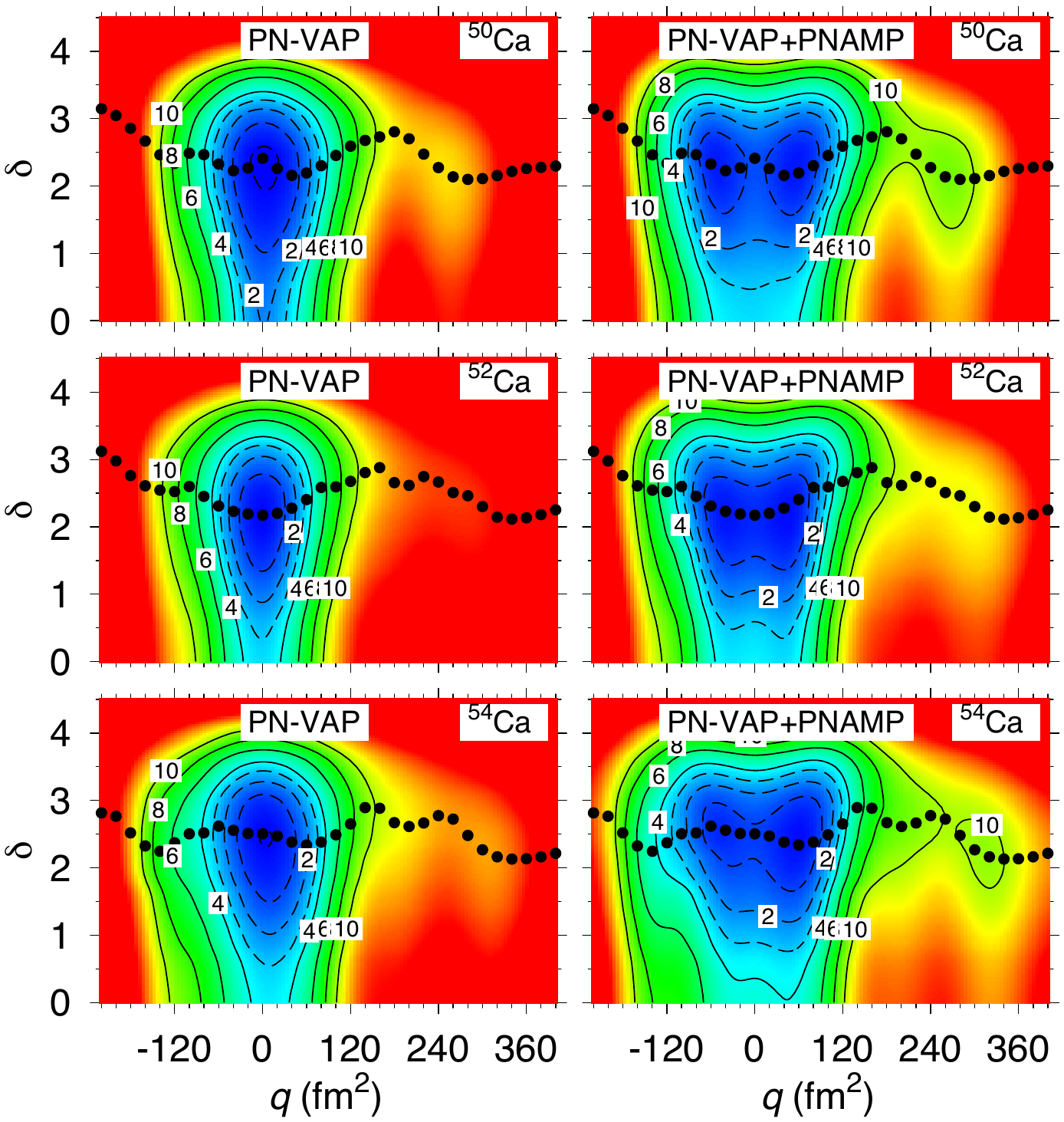}
\end{center}
\caption{Color online. Left panels: Potential energy surfaces for the nuclei$^{50}$Ca $^{52}$Ca and $^{54}$Ca in the PN-VAP approach. Right panels: Potential energy surfaces for the same nuclei  in the PN-VAP+PNAMP  approach for I=0. In both cases the contours follow the same interval as in Fig.~\ref{fig:pairfluc_Ti52}.}
\label{fig:pairfluc_Ca}
\end{figure}

Let us now turn to the Magnesium isotopes.  In Fig.~\ref{fig:pairfluc_Mg24_32} we display the potential energy surfaces for $^{24}$Mg and $^{32}$Mg just in the HFB and in the PN-VAP+PNAMP approach  for $I=0 \; \hbar$.  In the first row we find that $^{24}$Mg displays a stiff potential  in the HFB approach, it presents a structure of a deep prolate minimum $(q \approx 80$ fm$^2$) with  $\delta =0$ and a few MeV higher an oblate one ($q \approx -30$ fm$^2$).  We observe that this nucleus is more steep towards larger pairing correlations that the $^{52}$Ti.  In the PN-VAP+PNAMP case the prolate minimum shifts to $q \approx 100 $ fm$^2$ and $\delta \approx 2.0$ and the oblate one to $q \approx -40$ fm$^2$ and $\delta \approx 2.5$, the energy becoming even stiffer around the prolate minimum.
In the second row we display  $^{32}$Mg, in the HFB approach the energy minimum  has a spherical shape and  $\delta \approx 1.6$.   About 2 MeV higher  there is a prolate shoulder with $q \approx 80$ fm$^2$ and $\delta \approx 1.5$. In the PN-VAP+PNAMP approach, right panel,  we observe two deformed minima, the deepest one at $q \approx 90$ fm$^2$ and $\delta \approx 2.1$ and the oblate one at $q \approx -40$ fm$^2$ and $\delta \approx 2.5$,  about 2 MeV higher. The potential energy surface of the nucleus  $^{32}$Mg in the $q$ direction is wider and flatter than the one for  $^{24}$Mg and steeper in the $\delta$ values.

Finally  we represent in Fig.~\ref{fig:pairfluc_Ca} the same potential energy surfaces in the $q$, $\delta$ plane for the isotopes $^{50}$Ca, $^{52}$Ca and $^{54}$Ca, only for the PN-VAP and PN-VAP+PNAMP cases. Here we can study the evolution of the energy surfaces as a function of the neutron number. In the  PN-VAP approach (left panels)  all the three have a spherical ground state,  due to the Z=20 proton shell closure, the contours  are steep in the $q$ coordinate, but rather soft in the $\delta$ coordinate for  $\delta$ values less than about $3.5$.  Though one would
expect a softening in the $q$ direction with increasing neutron number  as an indication of the growth  in collectivity
we  observe that this is not the case for  $^{52}$Ca. As we will see below this is due to a subshell closure.
In the PN-VAP+PNAMP approach and  for $I=0\;\hbar$(right panels), we observe the well known effect of the angular momentum projection: the softening of the potential energy surface in the $q$ coordinate giving rise to  the  presence of  a structure of two minima, one prolate and one oblate at $-60$ fm$^{2}$ and $60$ fm$^{2}$ respectively. The softness in the $\delta$ coordinate remains however unchanged. 

\begin{figure*}[ht]
\begin{center}
\includegraphics[angle=-90 ,scale=0.6]{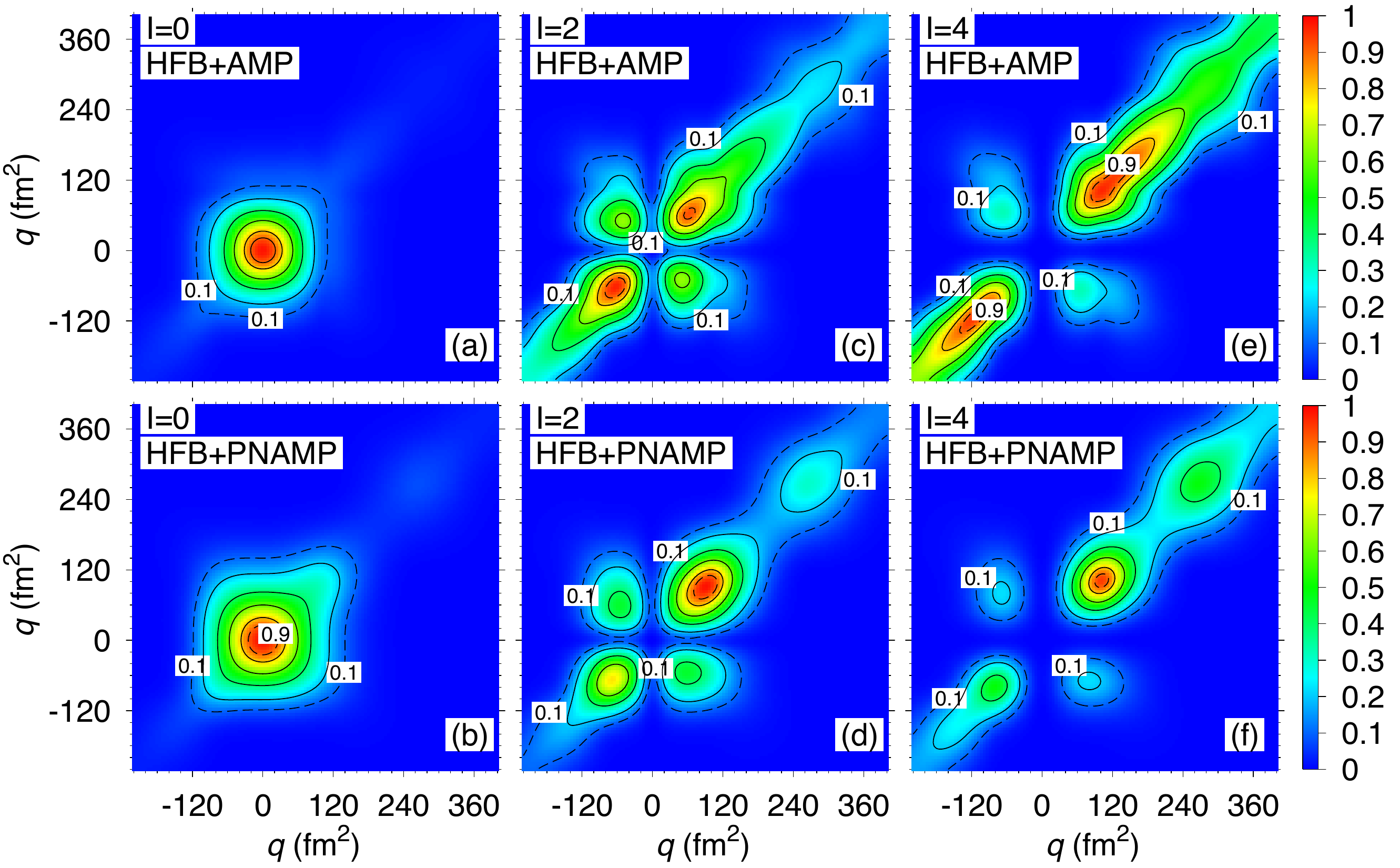}
\end{center}
\caption{Color online.  Renormalized  norm overlaps of the $^{52}$Ti nucleus. The original matrix elements have been divided  by a factor in order to have in each panel the maximum value equal to the unity. Top panels in the HFB+AMP approach, for  $I=0$ (factor=1.0), $I=2$ (factor=0.484) and $4$ (factor=0.316). Bottom panels in the HFB+PNAMP, for  $I=0$ (factor=0.191), $I=2$ (factor=0.218) and $4$ (factor=0.202). The continuous lines represent contours from 0.2 to 0.8 in steps of 0.2 and in dashed lines we depict the 0.1 contour and the 0.9 one. }
\label{fig:norma_1D}
\end{figure*}

  Other interesting issue is the fact that the 1D solutions of the PN-VAP approach are not only orthogonal to the contours of the corresponding 2D calculations, as they should,  but, to a  good approximation,  they are also orthogonal to the contours of the PN-VAP+PNAMP approach. This  indicates  that, at least for $I=0\; \hbar$, the PN-VAP 
1D  wave functions are, to a  good approximation,  self-consistent solution of the 2D PN-VAP+PNAMP approach.  One can check in Figs.~\ref{fig:pairfluc_Ti52} and \ref{fig:pairfluc_Mg24_32} that for those nuclei, and in this approach,   this is also the case. One can also realize  in  Fig.~\ref{fig:pairfluc_Ti52}, that to a lesser extend  this is also the case in the HFB case
as compared with the HFB+AMP and that this is not the case, at least around the energy minima, in the HFB+PNAMP approach. These facts are, as we will see later, a clear advantage of the PN-VAP intrinsic w.f.

The conclusion of this section is that,  in general and independently of the approach,  nuclei are soft in the pairing degree of freedom in the small to moderate strong pairing regime, however, for the very strong pairing the potential energy becomes very steep.

\section{Norm matrix elements} 
\label{normas}
In the diagonalization of the Hill-Wheeler equation, Eq.~\ref{HWeq}, enter two quantities, the Hamiltonian overlap and the norm
overlap. In the precedent section we have discussed in some extend the diagonal matrix elements of the Hamiltonian, we shall now concentrate on the
norm overlap to  get further inside of the different approximations. The main differences in the norm overlaps of the different  approaches appear between the PNP theories and the non-PNP ones.  Furthermore  in order to disentangle PNP effects from the ones stemming from VAP or PAV treatment we will
concentrate on the norms of the HFB+AMP approach and the HFB+PNAMP one. The norms of the PN-VAP+PNAMP are rather similar to the HFB+PNAMP ones and will not be shown here.

  Because of the huge amount of non-diagonal elements if the 2D  calculations, we will analyze in detail  the 1D case and some selected results in the 2D case. 

In  panel  a) of Fig.~\ref{fig:norma_1D} we display contour plots of the norm overlaps $\langle \phi(q)| P^{I=0}|\phi(q^{\prime})\rangle$  for the nucleus $^{52}$Ti in the  HFB+AMP approach  for $I=0$  as a function of $(q ,q^{\prime})$.  Since the HFB wave functions are normalized to the unity,  for the spherical case, i.e.,  for $q=q^{\prime}=0$, the norm overlap is just one.  Taking this point as center we observe a radial exponential decrease of the norm with  $q$.  For $q= 90$ fm$^2$ the norm is reduced to  one tenth of the value  at $q=0$ fm$^2$. In  panel b) of 
Fig.~\ref{fig:norma_1D} we now display $\langle \phi(q)| P^{I=0}P^N|\phi(q^{\prime})\rangle$, i.e., the norm matrix
in the HFB+PNAMP. For  $q=q^{\prime}=0$ this quantity is $\langle \phi(q)|P^N|\phi(q^{\prime})\rangle$ which
is not the unity, as a matter of fact it is $0.191$, because  the norm decreases as more projectors are added.
 In order to compare the different norms we have divided all panels with the corresponding factors  in order to have  the maximum value equal to the unity in all panels. The norm distribution in the HFB+PNAMP looks like the
one for the HFB+AMP though more extended and stretched along the diagonal. For the $I=2$ case, panels c) and d), we find that, as expected, the norms are zero around the spherical shape, they are large  around $|q|=|q^{\prime}| \approx 60$ fm$^2$ (for the HFB+PNAMP case at slightly larger values) and decrease smoothly along the diagonal and the perpendiculars to the diagonal ( $q+q^{\prime}=$ const.). In
the HFB+AMP case the distribution is practically symmetric with respect to the line $q = -q^{\prime}$, i.e., the prolate
and oblate parts do have similar values for the norm.
 In the HFB+PNAMP this is not the case, we find a clear predominance of the prolate side, this is due to the fact that
 this norm is sensitive to the pairing content of the wave functions.  For $I=4$ the situation goes along the same lines as the $I=2$ case but with stronger differences. Besides the oblate-prolate asymmetry present in the HFB+PNAMP approach we also observe that the HFB+AMP norms are more extended
 for $I=2$ and 4 than the HFB+PNAMP. This is a clear evidence of particle number mixing and of  spurious correlations  
in the HFB+AMP case.

\begin{figure}[h]
\begin{center}
\includegraphics[angle=270 ,scale=0.45]{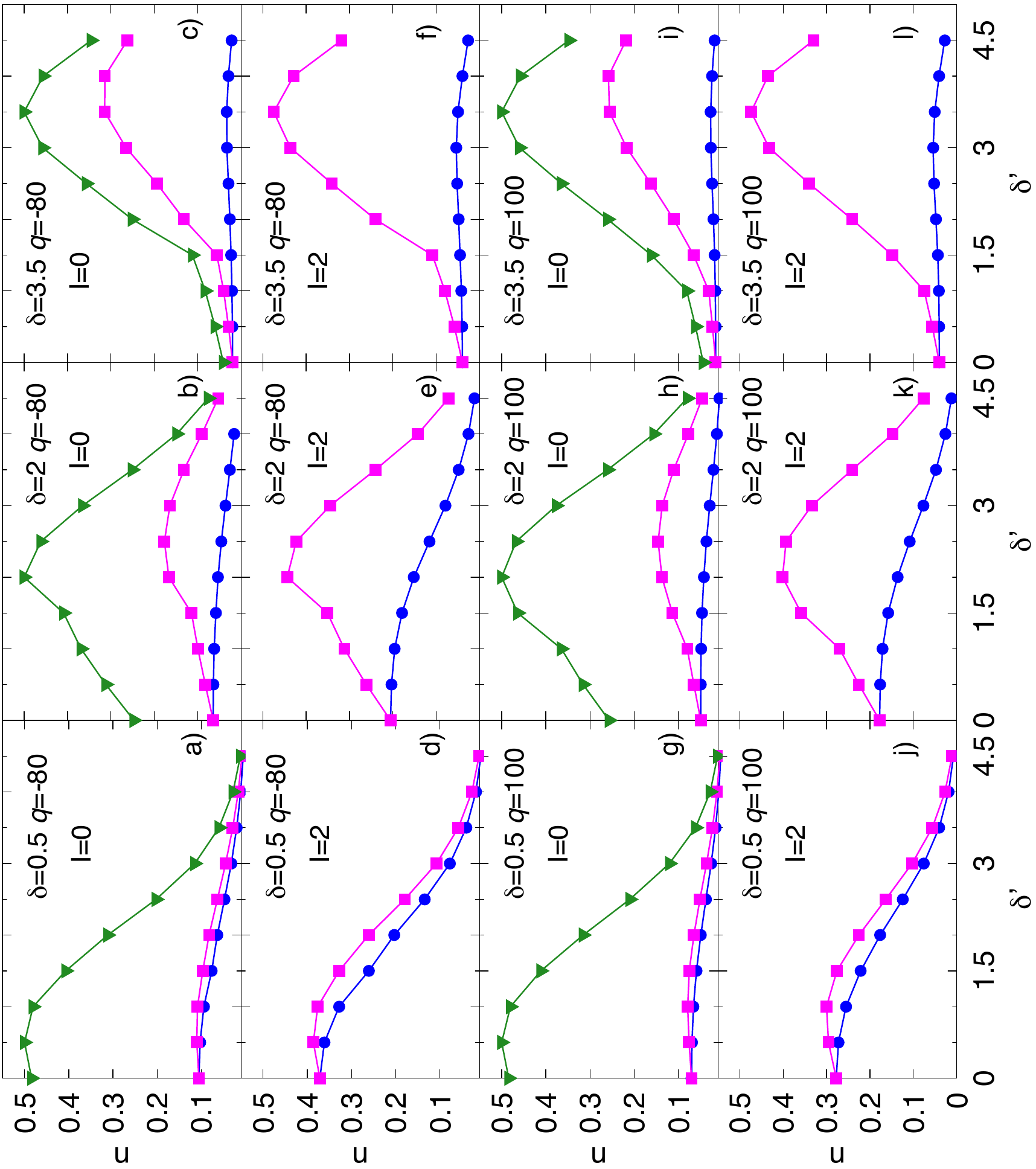}
\end{center}
\caption{(Color online) Norm matrix elements for $^{52}$Ti for different $q$ and $\delta$ values, see main text. In the HFB+PNAMP (blue bullets), HFB+AMP (magenta squares) and HFB (green triangles and only for $I=0 \hbar$) . Since the HFB w.f. are normalized to the unity these norms are multiplied by $0.5$  to make the pictures more legible.}
\label{fig:norma_2D}
\end{figure}

In Fig.~\ref{fig:norma_2D} we extend the analysis to the two dimensional case, i.e., we now
concentrate on the matrix elements $\langle \phi(q\delta)| P^{I}|\phi(q^{\prime}\delta^{\prime})\rangle$ (HFB+AMP approach) and $\langle \phi(q\delta)| P^{I}P^N|\phi(q^{\prime}\delta^{\prime})\rangle$  ( HFB+PNAMP approach) for $I=0$ and 2 and again for the nucleus $^{52}$Ti. We choose 
two $q$ deformations, namely, $-80$ fm$^2$ and  $100$ fm$^2$ which approximately correspond to the values of the energy minima, and three $\delta$ values which represent the weak $(\delta=0.5)$, the medium $(\delta=2.0)$ and the strong $(\delta=3.5)$ pairing regime~\footnote{ The terms weak, medium and strong pairing regime are  loosely used in reference to the domains where the diagonal matrix elements of the HFB and HFB+AMP peak.}.
 Furthermore we take $q^{\prime}=q$ in all the panels, the fixed $(q,\delta)$ values are given in each panel, the variable $\delta^{\prime}$ is indicated in the abscissa of each panel. We present calculations for $I=0$ (1st and 3rd  rows) and  $I=2$ (2nd and 4th rows). As a reference for  $I=0\; \hbar$ we also present the HFB overlaps 
$\langle \phi(q\delta)|\phi(q^{\prime}\delta^{\prime})\rangle$.  

While for fixed $q$ the HFB and the HFB+AMP approaches present a more or less Gaussian behavior around the diagonal matrix element $\delta^\prime=\delta$, the HFB+PNAMP does not.  For a  fixed $\delta$ value, the norm of this approach   either decreases or remains more or less constant for increasing $\delta^\prime$. The reason is very simple, if we expand the intrinsic w.f. in eigenstates  of the particle number operator
\begin{equation}
|\phi(q \delta^\prime \rangle = \sum_{\alpha^\prime N^\prime} C_{\alpha^\prime N^\prime}(q\delta^\prime) |\alpha^\prime, N^\prime\rangle,
\label{phi_exp}
\end{equation}
 the HFB+PNP norm is given by
\begin{equation}
\langle \phi(q \delta) | P^N|\phi(q \delta^\prime) \rangle = \sum_{\alpha} C_{\alpha N}(q\delta) C_{\alpha N}(q\delta^\prime).
\end{equation}

Obviously, for $\delta^{\prime}=0$, $\sum_{\alpha}|C_{\alpha N}(q\delta^\prime)|^2=1$.  For large $\delta^\prime$ the spread in the number of particles grows considerably and  more   terms contribute to the expansion (\ref{phi_exp}).  But since
\begin{equation}
\sum_{\alpha^{\prime} N^{\prime}} |C_{\alpha^{\prime}N^{\prime}}(q\delta^\prime)|^2 =1,
\end{equation}
for a given $N$,   $\sum_{\alpha}C_{\alpha N}(q \delta)C_{\alpha N}(q \delta^\prime)$ will in general decrease with growing $\delta^\prime$. In this argument we have omitted the angular momentum operator which could modulate 
the behavior of the norm.

If we now compare both approaches the tendency is 
clear: in the small pairing regime, left panels,  there are little differences between both approaches independently of the $q$ value, in the medium pairing regime, medium panels, we find larger deviations and in the strong regime, right panels, even
larger.  In the strong pairing regime the HFB+AMP curves for  $I=0\; \hbar$ look somewhat like the HFB one. Furthermore, we observe that the deviations of the HFB+AMP with respect to the HFB+PNAMP
increase with the angular momentum.  An interesting observation is the fact that whereas the 
HFB+PNAMP approach, in general, presents  a  mostly decreasing behavior with increasing $\delta^{\prime}$,   the HFB+AMP one coincides with the HFB+PNAMP one at the small $\delta^{\prime}$ then increases with growing $\delta^{\prime}$ up to a maximum value for $\delta^\prime =\delta$, then decreases again and in the limit
 of very large $\delta^{\prime}$ would coincide with the HFB+PNAMP value. The last one is clearly
 the semi-classic limit.  Unfortunately,  large deviations take place between both approaches at the  $\delta^{\prime}$ values which are the commonest in nuclei.

\section{Spectra}
\label{espectros}
 
We  discuss now the results of the SCCM calculations, for which the HW (Eq.~\ref{HWeq})  has to be solved. Before discussing  the excitation spectra we will comment on the limitations of our approaches. In our description we are considering mainly collective degrees of freedom, namely the quadrupole deformation and the pairing gap. Though we are considering different nuclear shapes and, in principle, single particle degrees of freedom can be expanded as  linear combinations of different configurations, we cannot claim to describe properly genuine single particle states  but only in an approximate way. Collective states on the other hand are very well described in our approach.

The HW equation has to be solved separately for each value of the angular momentum, the diagonalization of this matrix provides the yrast and the excited states,  $I^{+}_{1}, I^{+}_{2}, I^{+}_{3},...$ for each angular momentum. These energy levels, normalized to the  ground state energy ($0^{+}_{1}$), 
provide the spectrum of the nucleus.  Again, we will study the three cases we are focused on, namely HFB+AMP, HFB+PNAMP and PN-VAP+PNAMP. In order to evaluate the impact of the pairing fluctuations on the different observable we  consider the solutions of the HW equation in 1D, with one coordinate $(q)$,  and in 2D, with two coordinates $(q, \delta)$. We have calculated the four lowest states for each angular momentum.

    Though we are acquainted with the diagonal matrix elements of the HW equation through the energy surface contour plots previously discussed,  the magnitude of the non-diagonal elements and thereby the energy gain after the solution of the HW equation depends on the approach. In general, because of self-consistency and due to the quality of the approach before the HW diagonalization, we expect a smaller energy gain -as compared with the energy minimum of the potential energy-  in the PN-VAP+PNAMP case than in the HFB+PNAMP one.  For the special case of $I=0\;\hbar$ we can make a thorough discussion of the different contributions because in this case we can  solve the HW equation within our framework even for the plain HFB approach. The reason is  that in a {\it semiclassical} approach to eigenstates of the angular momentum  (Cranking approach),  one has to add the term $-\omega J_{x}$ to the variational principle Eq.~\ref{E_prime}. The cranking frequency being determined by  the condition $\langle J_{x}\rangle =  \sqrt{I(I + 1)}$. Since our HFB wave functions do not break time reversal, the condition 
 $\langle J_{x}\rangle = 0$ is always satisfied and the plain HFB approach can be considered to approximately describe $I=0\;\hbar$ states.

 \begin{figure}[h]
\begin {center}
\includegraphics[angle=0, scale=0.4]{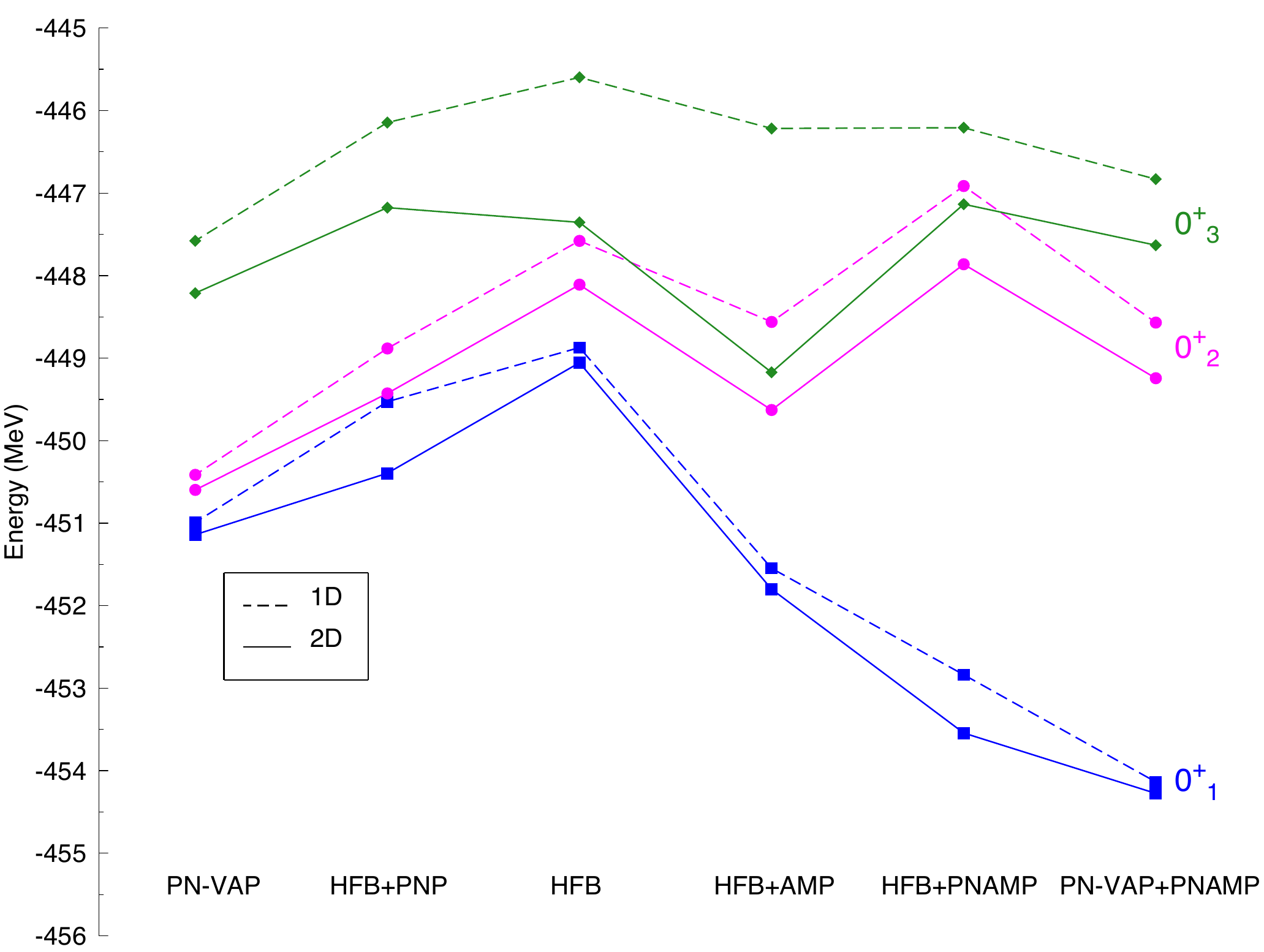}
\caption{(Color online) Evolution of the total energies of the ground state energy $0^{+}_{1}$ (blue) and excited energies $0^{+}_{2}$ (magenta) and $0^{+}_{3}$ (green) as a function of the different approximations for $^{52}$Ti. Dashed lines  correspond to 1D calculations and continuous are for 2D calculations.}
\label{fig:abs_spec_Ti}
\end{center}
\end{figure}

 \begin{table}[tbh]
\begin{center}
\begin{tabular}{|c|c|c|c|c|}\cline{3-5}
\multicolumn{2}{c|}{} & $0^+_{1}$ & $0^+_{2}$ &  $0^+_{3}$ \\ \hline 
PN-VAP   & 1D & -450.991 & -450.415  & -447.579 \\ \cline {2-5} & 2D & -451.140 & -450.598  & -448.213\\ \cline {2-5} \hline 
HFB+PNP    & 1D & -449.528 & -448.883 & -446.145  \\ \cline {2-5} & 2D & -450.395 & -449.428 & -447.176 \\ \cline {2-5}  \hline 
HFB        & 1D & -448.872 & -447.578 & -445.598   \\ \cline {2-5} & 2D & -449.056 & -448.111 & -447.355 \\ \cline {2-5}  \hline 
HFB+AMP    & 1D & -451.547 & -448.561 &	-446.217\\ \cline {2-5} & 2D & -451.800 & -449.629 & -449.173 \\ \cline {2-5} \hline 
HFB+PNAMP      & 1D  & -452.837 & -446.915 & -446.208 \\ \cline {2-5} & 2D  & -453.543 & -447.863 & -447.135\\ \cline {2-5} \hline 
PN-VAP+PNAMP   & 1D  & -454.136 & -448.570 & -446.832 \\ \cline {2-5} & 2D  & -454.275 & -449.245 & -447.632 \\ \cline {2-5} \hline 
\end{tabular}
\end{center}
\caption{ Absolute energies of the three lowest  states in several approaches for ${^{52}}$Ti, in MeV}
\label{Table1}
\end{table}

In Fig.~\ref{fig:abs_spec_Ti}  and in Table~\ref{Table1} we display the absolute energy of the three first eigenvalues of the HW equation for $I=0\;\hbar$ in different approaches and in the 1D and 2D cases. In the abscissa  the different approaches are indicated, the line connecting them is used to guide the eye. The simplest approximation is provided by the solution of the HW equation with plain HFB w.f. without any kind of projectors. The other ones correspond to  adding more  complexity to the  wave functions. The general
behavior is the following:  1) For a given approach the 2D approach is always lower than the 1D one. This is clear since within a given approach going from 1D to 2D means to add an extra degree of freedom. 2)  The ground state energy always decreases with increasing complexity of the wave function.
If we now concentrate on the HFB states we observe that  by a separate PNP (left of the  HFB solution) or an AMP  (right) one  obtains,  in general,  an energy decrease, however, when a simultaneous PNAMP is performed an interference effect appears (see states $0^{+}_{2}$ and $0^{+}_{3}$  in 1D and 2D) and as a consequence some states increase its energy as compared with the simpler approach. We also  observe that, as expected, in the VAP approaches (PN-VAP and PN-VAP+PNAMP) the energy gain from 1D to 2D is smaller than in the 
HFB approaches (HFB+PNP and HBF+PNAMP). We can also appreciate in the two most left (right) approaches the difference between a VAP and a PAV approach for the particle number case and for a given 1D or 2D case. For the ground state and in the 1D case the energy gain of a VAP as compared to a PAV approach  amounts to $1.5$ MeV, the same quantity but now with AMP amounts to $1.3$ MeV.
 Finally, we mention that the total energy gain of the $0^{+}_{1}$ state from the simplest HFB(1D) description to the most sophisticated one PN-VAP+PNAMP(2D) amounts to $5.4$ MeV.
 
\begin{figure*}[tbh]
\begin {center}
\includegraphics[angle=-90, scale=0.45]{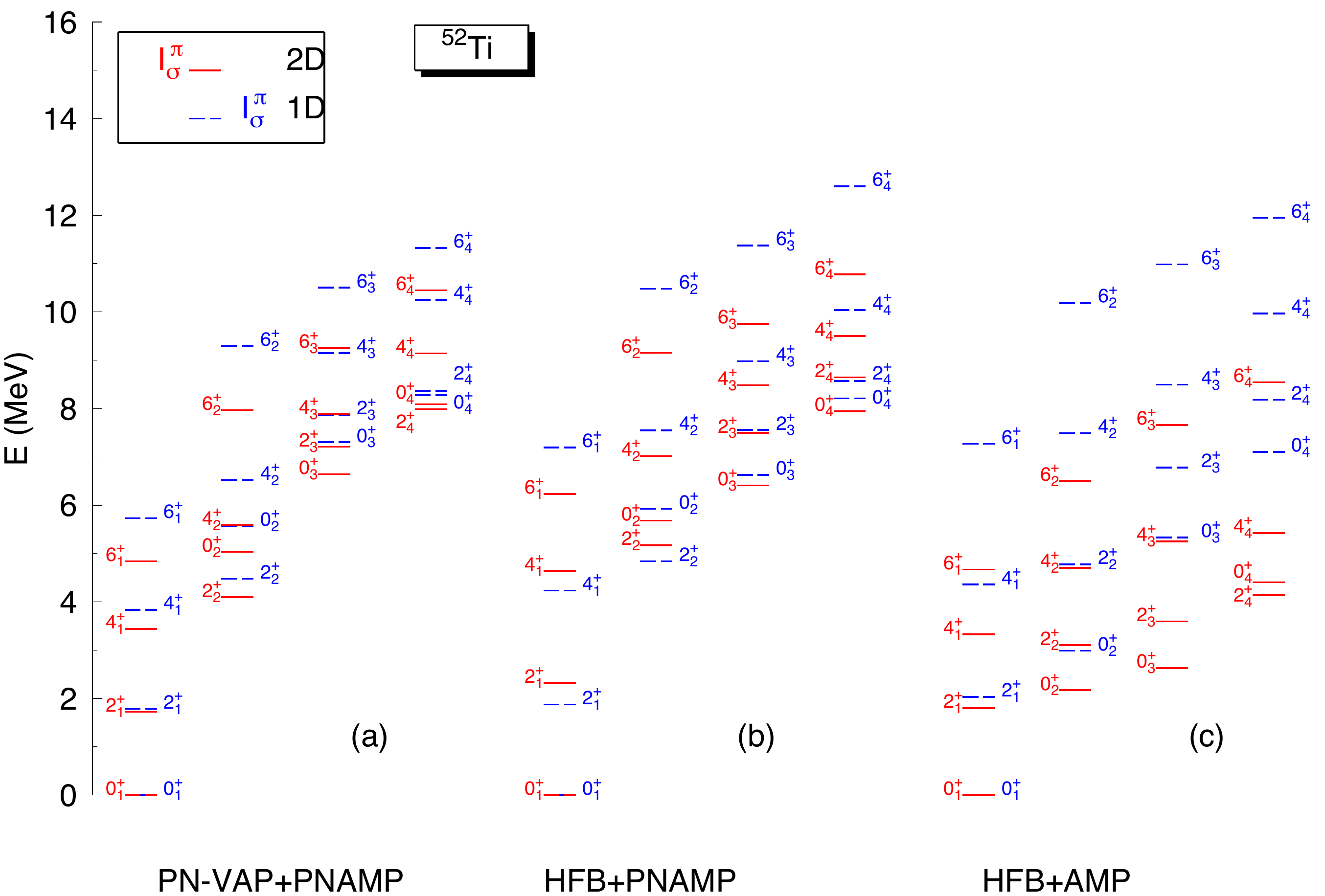}
\caption{(Color online) Spectra of $^{52}$Ti in the PN-VAP+PNAMP (left), HFB+PNAMP (middle) and HFB+AMP (right) approaches.  The four lowest states for spin $0^{+}, 2^{+}, 4^{+}$ and $6^{+}$ are represented in the 1D (dashed lines) and 2D(continuous lines).}
\label{fig:spect_Ti}
\end{center}
\end{figure*}

We now look for the excitation spectra, but before making a detailed description let us just mention a very general argument to guide our discussion. 
The comment above on the cranking approximation can also be interpreted in the light of a  {\it quantum} approximation to an  angular momentum VAP method. According to  the  Kamlah expansion \cite{Kamlah}  a VAP of the angular momentum  can be approximated, to first order, in the following way:  the intrinsic HFB wave function, $|\phi \rangle $,  is determined by minimizing the energy $E^\prime= \langle \phi | \hat{H} | \phi \rangle-\omega \langle \phi | \hat{J}_x | \phi \rangle$ with $\omega$ determined by the
 constraint $\langle J_{x}\rangle =  \sqrt{I(I + 1)}$, the energy is provided by $E^I= \langle \phi | \hat{H} P^I | \phi \rangle/\langle \phi | P^I | \phi \rangle$.  Since for $I=0\;\hbar$,  $\langle \phi | \hat{J}_x | \phi \rangle=0$, the Kamlah prescription do apply in this case  in the three approaches,  but for $I\neq 0\;\hbar$ this is not the case because our w.f. do not break time 
 reversal and they cannot fulfill the constraint on the angular momentum.  That means that our approaches
 favor the states with $I=0\;\hbar$ because for them an approximate VAP for the angular momentum is performed.  For 
 $I\neq 0\;\hbar$ this is not the case and  we just do plain PAV.  From this arguments and from this perspective it is obvious that,  the quality of the approach diminishes with growing $I$-values.
 That means,  the relative energy gain will be largest for $I=0\;\hbar$,  and for $I\neq 0\;\hbar$  it will comparatively decrease with increasing $I$. Thus in our current approach we  predict stretched spectra, this will not be the case anymore if we break the time reversal symmetry \cite{zdudoba}. 
 
In Fig.~\ref{fig:spect_Ti} we present the excitation spectrum for $^{52}$Ti  in our three basic  approaches and in the 1D and 2D calculations. The levels are ordered just by the energy. 
In the left hand part we display the most complete  approach, namely the PN-VAP+PNAMP. 
The general trend is that the 1D calculation is more stretched that the 2D one.  This is a clear manifestation of the following fact: Since the 1D and the 2D calculation are self-consistent the ground state energy {\em before} the HW diagonalization, i.e., the minimum of the potential energy surfaces, is the same in both calculations and even after the HW diagonalization they are rather similar, see Table \ref{Table1}.  This result is a consequence of the fact that the variational principle used to determine the wave functions $|\phi\rangle$ favors ground states.  
In the 1D calculations there is no room for the excited states to change the pairing content of a given w.f., however,  in the 2D calculations the flatness of the pairing degree of freedom opens the possibility of choosing  different pairing energies for a given deformation $q$ allowing thereby  an energy lowering.  We see, therefore  that the consideration of additional degrees of freedom partially {\it compensates} the above mentioned problem of approximate VAP for  $I=0\;\hbar$ versus PAV for $I\neq 0\;\hbar$. In reality we are doing a restricted VAP, see Ref.~\cite{PhysRevC.71.044313} for more details.

In the middle of Fig.~\ref{fig:spect_Ti} the HFB+PNAMP spectrum is presented. This spectrum is, in general, more stretched than the PN-VAP+PNAMP one. Another difference is the fact that the ordering of some levels,
in particular the Yrast ones, of the 1D and 2D calculations are inverted as compared with the PN-VAP+PNAMP one.  The reason for this behavior is the lack of self-consistency  (in the sense of the end of Sect.~\ref{PES_2D})  of this approach.  As we can see in the panel (e) of Fig.~\ref {fig:pairfluc_Ti52} the path of the 1D solution in the $(\delta,q)$ plane, i.e. the bullets line,  goes along lines of smaller pairing correlations than the minima displayed by the 2D contour plots. Consequently, in 1D  the mass parameter associated with the collective motion is  larger than in 2D and the associated  spectrum more compressed in the former than in the latter one. This effect combined with the additional degree of freedom of the 2D discussed above makes that  only the lower levels are inverted. 

Finally in the right part of Fig.~\ref{fig:spect_Ti} the HFB+AMP approach is displayed. First, we observe  very much compressed spectra  as compared with the other approaches.
 It is remarkable  the fact that all states with the same spin are  much closer to each other than in the PNP approaches. One furthermore notices the unusual large lowering of the 2D states as compared with the 1D ones.  These facts seems to indicate that, as mentioned in Sect.~\ref{normas},  there is too much mixing in the solution of the HW equation due to spurious contributions stemming from the non-conservation of the particle number symmetry. One also observes that contrary to the inversion of the HFB+PNAMP, the inversion of the 1D and 2D levels does not take place in this case.  This is due to the fact that in this case we are  more self-consistent than in the HFB+PNAMP case.
 
 \begin{figure}[tbh]
 \begin{center}
\includegraphics[angle=0, scale=0.5]{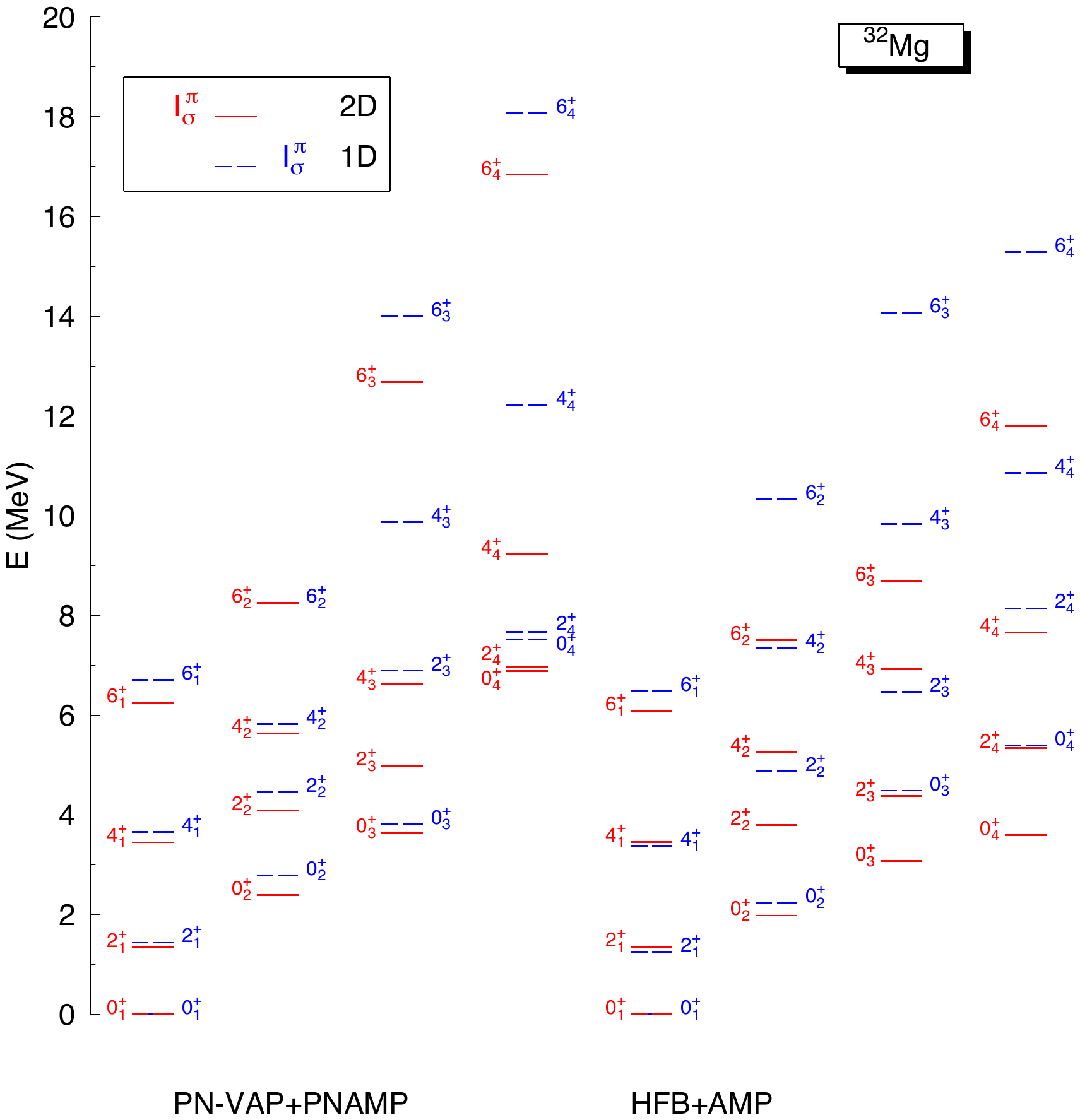}
\caption{ (Color online)Spectra of $^{32}$Mg in the PN-VAP+PNAMP (left) and the HFB+AMP (right).}
\label{fig:spect_Mg32}
\end {center}
\end{figure}

\begin{figure*}[tbh]
\begin{center}
\includegraphics[angle=-90, scale=0.4]{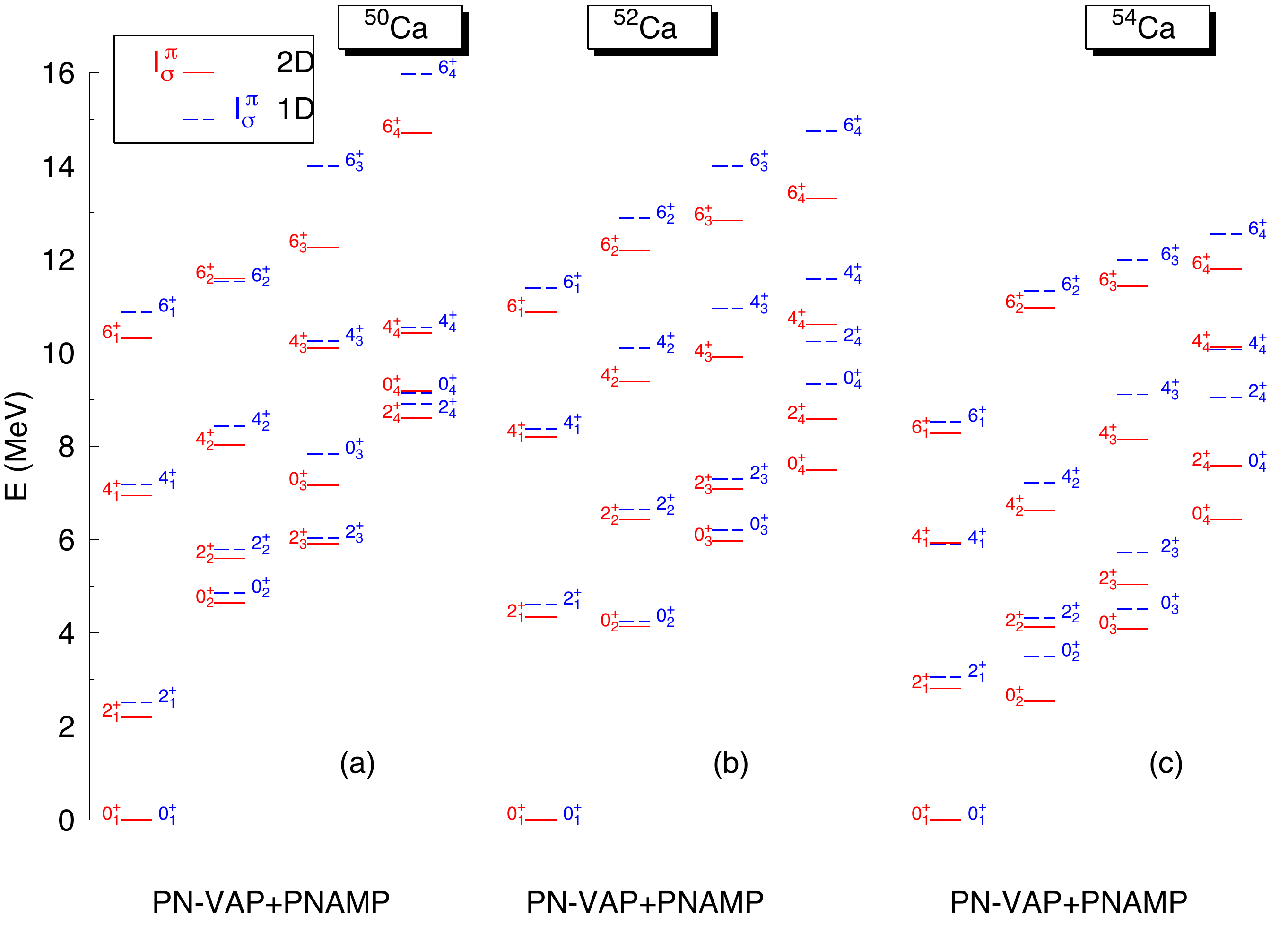}
\end {center}
\caption{(Color Online) Spectra of $^{50}$Ca, $^{52}$Ca $^{52}$Ca in PN-VAP+PNAMP approach.}
\label{fig:spect_Ca}
\end{figure*}

 Concerning the 2D spectra in the three approaches one can understand the degree
 of compression of the spectra by looking at the right hand panels of 
 Fig.~\ref{fig:pairfluc_Ti52}. We observe that by far the softest surface towards  small pairing correlations is the HFB+AMP,  then, though to a lesser extend,  PN-VAP+PNAMP (in the energetic relevant part, i.e., around the minima)  and finally HFB+PNAMP relatively close to the former one.  Correspondingly we expect the HFB+AMP spectrum to be the most compressed,  followed by  PN-VAP+PNAMP and finally HFB+PNAMP relatively close to the latter one.

In Fig.~\ref{fig:spect_Mg32} we display another example of HFB+AMP spectrum versus the PN-VAP+PNAMP one, this time for the deformed collective nucleus $^{32}$Mg. Though clear  differences are observed between both spectra, specially for the $I_3$ and $I_4$ states,  in the  $I_2$ states the difference, at least for the lowest ones, is not as large as for the Titanium case. In the PN-VAP+PNAMP case the whole spectrum is compressed because $^{32}$Mg is very collective as on can see in the broad potential displayed in  Fig.~\ref{fig:pairfluc_Mg24_32} and rather steep in the pairing fluctuations. For  the nucleus $^{24}$Mg (not shown here), on the other hand,  we find large differences between both approaches. Of course, in the latter one and in the HFB approach the pairing correlations  vanish at the potential minimum, see Fig.~\ref{fig:pairfluc_Mg24_32}, inducing a large moment
of inertia and the corresponding compression of the HFB+AMP spectrum.  In this case is difficult to disentangle 
both effects, i.e., the differences caused by the smaller pairing correlations (as compared with the  PN-VAP) and   
the spurious mixing caused by the non conservation of the particle number.

Lastly, we present in Fig.~\ref{fig:spect_Ca} the spectra of the isotopes $^{50-52-54}$Ca in the PN-VAP+PNAMP to see the evolution of the spectra with the neutron number. We observe  shifts in the 2D calculations with respect to the 1D with the same tendencies as the nuclei discussed above. In $^{50}$Ca we observe an inversion of the $0^{+}_3$ and $0^{+}_4$ levels with the respective $2^{+}_3$ and $2^{+}_4$ levels that does not show up in $^{52-54}$Ca.   In principle, one would expect an increase in collectivity with growing  neutron number. Looking at the spectra we find that this not the case, the nucleus $^{52}$Ca
does not appears as a smooth interpolation of $^{50}$Ca and $^{54}$Ca, as a matter of fact, the low lying states of $^{52}$Ca are higher in energy than in  $^{50-54}$Ca.  We observe, in particular,  that the $2^{+}_{1}$  state of the nucleus $^{52}$Ca is higher in energy than in its neighbors, this fact has been interpreted   as a  sub-shell closure at $N=32$. The discussions  going on \cite{PhysRevLett.87.082502, Prisciandaro200117, Rodriguez2007} about the hypothetical shell closures at $N=32$ and $N=34$ has been settle by a recent
measurement of the excitation energy of the $2^+_1$ level in $^{54}$Ca \cite{steppenbeck_nature}
Our prediction for this state taking the pairing degree of freedom into account is in agreement   with the experimental finding, see also \cite{Rodriguez2007}.  

We conclude this section by stressing the impact of the pairing of the nuclear spectra and the relevance of the
particle number projection to avoid unwanted mixing.

\section {Collective wave functions}
\label{coll_w_f}
In this section we discuss the collective wave functions, see Eq.~\ref{Coll_wf},  solution of the Hill-Wheeler equations in one, 
$g(q)$, and
two dimensions,  $g(q,\delta)$,  and in the three basic approaches only for the nucleus $^{52}$Ti.
 \begin{figure}[h]
\begin{center}
\includegraphics[angle=0 ,scale=0.62]{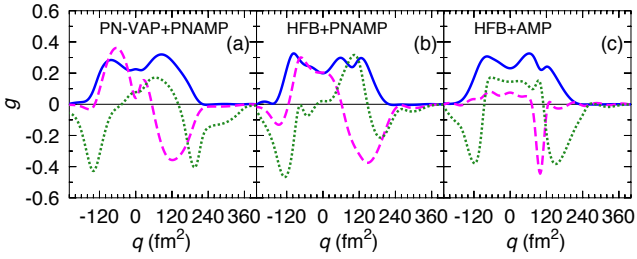}
\end{center}
\caption{(Color Online) The wave functions of the lowest $0^+$ collective states for the nucleus $^{52}$Ti in various approaches: Left in the PN-VAP+PNAMP, middle in the HFB+PNAMP   and right in the HFB+AMP approach in the 1D calculations. 
$0^+_1$ continuous blue line, $0^+_2$  dashed magenta line and $0^+_3$  dotted green line. The w.f. of the  $0^+_2$ state in the HFB+AMP approach has been multiplied by a factor 0.5.}
\label{fig:WF_1D_Ti_signo}
\end{figure}

\begin{figure*}[tbh]
\begin{center}
\includegraphics[angle=-90,scale=0.5]{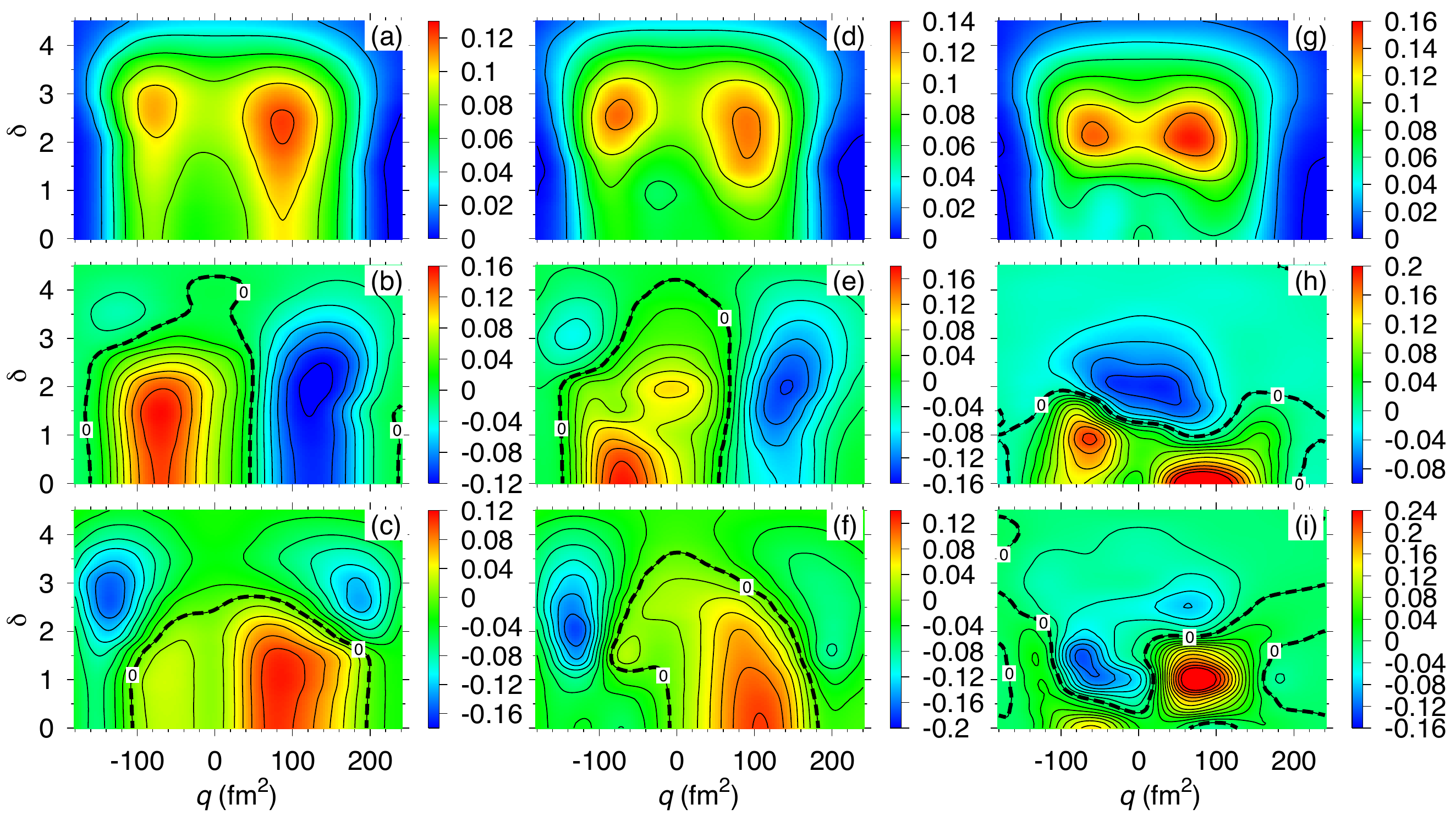}%
\end{center}
\caption{ (Color online) Contour lines of the wave functions of the three lowest $0^+$ states of $^{52}$Ti in different approaches in the 2D calculations.  The step size is 0.02, for the states. The thick dashed lines correspond to the zeros of the wave function.  To get larger resolution the x-axis runs from -180 fm$^2$ up to 240 fm$^2$ at variance with former figures.}
\label{fig:WF_J0_52Ti}
\end{figure*}

In Fig.~\ref{fig:WF_1D_Ti_signo} we present the  HW wave functions of the $^{52}$Ti nucleus in the one-dimensional
case in the three basic approaches. The corresponding potential  energy curves have been plotted in
Fig.~\ref{fig:1D}.  In the left top panel of  Fig.~\ref{fig:1D} the PN-VAP+PNAMP 
 potential energy curve display two quasi-coexistent minima the lowest one prolate and the other one oblate, consequently the w.f.'s  (see Fig.~\ref{fig:WF_1D_Ti_signo})(a) of the  $0^{+}_{1}$ and $0^{+}_{2}$ states display a two hump structure with maxima (or maximum and minimum) at these values, the $0^{+}_{2}$ with a node as one would expect for a $\beta$ vibration.
 The $0^{+}_{3}$ state, on the other hand,
peaks  at large deformations in the prolate and the oblate potential shoulders  and it has a two nodes structure.   In the HFB+PNAMP approach, the potential energy curve ( see Fig.~\ref{fig:1D}) presents also two minima although not as pronounced as in the previous case and somewhat narrower. The kink around 100 fm$^2$, has its origin in the neutron pairing energy collapse of the HFB wave function at this point, see right top panel of Fig.~\ref{fig:1D}.
 and it causes the split of the prolate bump of the  $0^{+}_{1}$ state.  The 
w.f. of the $0^{+}_{2}$ state, similarly to the PN-VAP+PNAMP case presents a two bump structure.   The $0^{+}_{3}$ one, though similar to the PN-VAP+PNAMP too,  shows a decrease of the prolate shoulder,  in part due to the
collapse of the pairing energy of the HFB w.f.  at this point,  and to the fact that the magnitude of the shoulder in the potential energy is smaller than in the former case.  As it is well known at the energy minima the level density decreases and so does the pairing energy in the HFB approach  (in some cases it even collapses !), in a PN-VAP approach, obviously, this is not the case.   In the HFB+AMP
case the potential energy has the same two minima structure and it is even narrower than before.  The wave function of the $0^{+}_{1}$ state has the right hand side of the split bump smaller as compared with  the former case. The big 
difference appears in the $0^{+}_{2}$ state where the w.f. rises at the place where the neutron pairing gap gets zero. The $0^{+}_{3}$ state, on the other hand, peaks at  smaller prolate $q$ values  due to the almost disappearance of the shoulder and the fact that the potential energy is somewhat narrower.

Concerning the $(q,\delta)$ calculations,  the potential energy surfaces have been already discussed in Fig.~\ref{fig:pairfluc_Ti52},   the two dimensional wave functions are presented in Fig.~\ref{fig:WF_J0_52Ti}. We start again with the PN-VAP+PNAMP case. In panel (a) the contours lines of the wave function of the $0^{+}_{1}$ state are shown.
In strong correspondence with the lowest right panel  of  Fig.~\ref{fig:pairfluc_Ti52} it presents a two bump structure, rather soft in the pairing degree of freedom, with a predomination of the prolate side. The bumps maxima are located at $q$-values  close to the 1D case and centered at $\delta$ values close to the self-consistent solution (see bullets in Fig.~\ref{fig:pairfluc_Ti52}). The $0^{+}_{2}$ state, panel (b), displays also   a two bump structure, this time with the maximum in the oblate side and soft in $\delta$. The maxima are located at $\delta$ values smaller than the $0^{+}_{1}$ state. It presents a nodal line at
$q\approx 50$ fm$^2$ as correspond to a $\beta$ vibration in two dimensions. 
  The $0^{+}_{3}$ state, panel (c), presents a three-peak structure, two at large deformations and large pairing correlations and a smaller one around 80 fm$^2$ with smaller pairing correlations. This situation is similar to the 1D case where at similar $q$-values the same peaks are found. The fact that the large deformations peaks do have
  strong pairing correlations is due to the fact that the level density is very high at these
  deformations and that the 2D calculations allow that  a given $q$ value can take different
  pairing content for different collective states.   
  
   Looking at  panel (e) of Fig.~\ref{fig:pairfluc_Ti52}, the 1D plots  of Fig.~\ref{fig:WF_1D_Ti_signo} and taking into account the discussion above one can interpret the 2D  wave functions of the HFB+PNAMP approach very easily.
The main difference with the former case is that the beta vibration and the $0^+_3$ state in this case are not as pure as in the PN-VAP+PNAMP case.

The HFB+AMP collective wave functions look more different than the ones of the two former
approximations. The $0^{+}_{1}$ one still maintains the two bump structure though more pronounced than in the  HFB+PNAMP case. The two peaks  are located at smaller $\delta$ values and closer to each other in the $q$ coordinate.  The strong peak of $0^{+}_{2}$ at $q\approx 110$ fm$^2$ of the 1D case is also present in the 2D one but now a second peak appears at $q\approx -60$ fm$^2$. The $0^{+}_{3}$  wave function reminds the 1D in some aspects. One has the impression that the
$0^+_2$ and $0^+_3$ states  of the 1D and of the other 2D PNP approaches get mixed in the HFB+AMP approach.
 In this approach the wave functions are more concentrated than in the two former ones. One can quantify this effect noticing that
 the maxima (in absolute value) of the w.f. in the PN-VAP+PNAMP are 0.125, 0.158 and 0.198
 for the $0^{+}_{1}$, $0^{+}_{2}$ and $0^{+}_{3}$ respectively.  For the HFB+PNAMP
0.133, 0.183 and 0.295 in the same order, and  for the HFB+AMP, 0.161, 0.200 and 0.357.
 The trend is clear, the concentration of the wave function increases as the quality of
 the respective approach decreases.

Interestingly, though the potential energy surfaces in the three cases are rather  similar, see panels (b), (e)  and (f) of Fig.~\ref{fig:pairfluc_Ti52}, the wave functions of the HFB+AMP are rather different from the other ones. This has obviously  to do  with the  non-diagonal elements of the Hamilton overlap and the norm overlap. The former through the dynamical corrections and the latter through the linear dependence of the basis states.

We conclude this section again underlining the relevance of the particle number projection for a proper description of the 
properties of atomic nuclei.

\section{Pairing vibrations}
\label{pair_vib}

One could also rise the question about the existence of pure pairing vibrations. They are on their own very interesting and a simultaneous study of the shape and pairing fluctuations will allow us to disentangle if there exist genuine pairing vibrations  or they are washed out because of the energy predominance of the quadrupole ones.   In this section we discuss this issue in the framework of the PN-VAP+PNAMP approach.

\begin{figure}[h]
\begin{center}
\includegraphics[angle=0, scale=0.5]{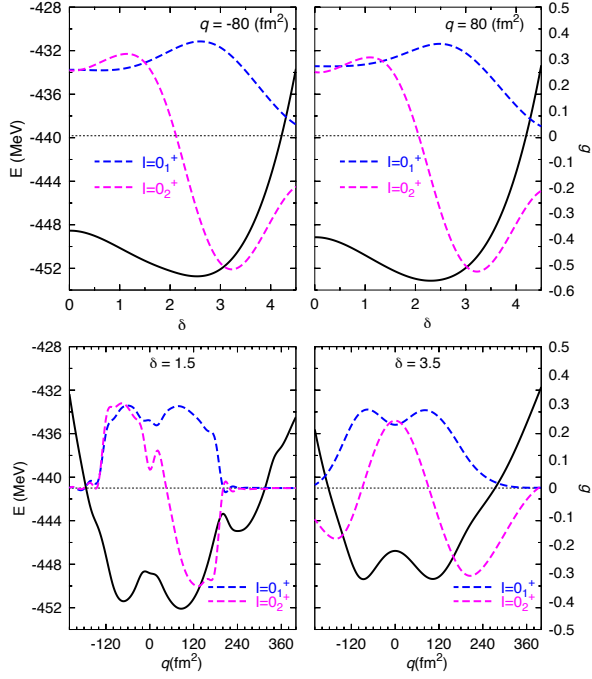}
\caption{(Color Online) Potential well and wave functions of the two lowest states in 1D calculations for the nucleus $^{52}$Ti. Top: For fixed $q$-value and with $\delta$ as generator coordinate, left for $q=-80$ fm$^2$ and right for $q=+80$ fm$^2$. Bottom: For fixed $\delta$-value and with $q$ as generator coordinate, left for $\delta=1.5$  and right for $\delta=3.5$}
\label{pai_vib_1}
\end{center}
\end{figure}
Pairing vibrations appear naturally when 1D calculations are performed using $\delta$ as the generating coordinate and fixing the deformation.  In the top panels of Fig.~\ref{pai_vib_1}, we show the potential energy (in black) and the collective wave functions, Eq.~\ref{Coll_wf},  for the states $0^{+}_{1}$ (blue) and $0^{+}_{2}$ (magenta)  as a function of $\delta$ at a fixed value of $q=-80$ ${\rm fm}^2$  and  $q=80$ ${\rm fm}^2$  for the nucleus $^{52}$Ti.  The ground state wave function does not present any node while the first excited state, $0^{+}_{2}$, shows one node as corresponds to a pairing vibration.

We now consider simultaneously the pairing and the quadrupole deformation degrees of freedom.  The corresponding potential energy contour plot has been already discussed  in  Fig.~\ref{fig:pairfluc_Ti52}. Since we expect vibrations associated to the deformation ($q$) and with the pairing ($\delta$) we now have to consider at least  three states, the ground and the two lowest excited states, to seek for the shape and the pairing vibrations.  In the left panels
 of Fig.~\ref{fig:WF_J0_52Ti} we have shown the wave functions, Eq.~\ref{Coll_wf}, of the lowest three states.  Since we have two coordinates we expect that the
nodes of the wave functions of one dimensional case will now turn to  nodal lines. The ground state wave functions, see panel (a),  is everywhere positive, it has two maxima in a direct correspondence with the two minima of the potential well. The first excited state presents a maximum at $q \approx -60$ fm$^2$  and a minimum at $q \approx +140$ fm$^2$ with  a nodal line  in between at $ q \approx 50 $ fm$^2$. This state is clearly identified as a beta vibration.
In panel (c) the second excited state is depicted. It has two strong minima at large $q$ values and a maximum at
$q \approx 80$ fm$^2$ with a long tail towards the oblate side. The node line in this case is formed by two segments
perpendicular to the $q$ axis and a curved segment more or less perpendicular to the $\delta$ axis. The two first
segments  would be an indication of a two phonon beta vibration and the latter one of a pairing vibration.

\begin{figure}[h]
\begin{center}
\includegraphics[angle=0, scale=0.6]{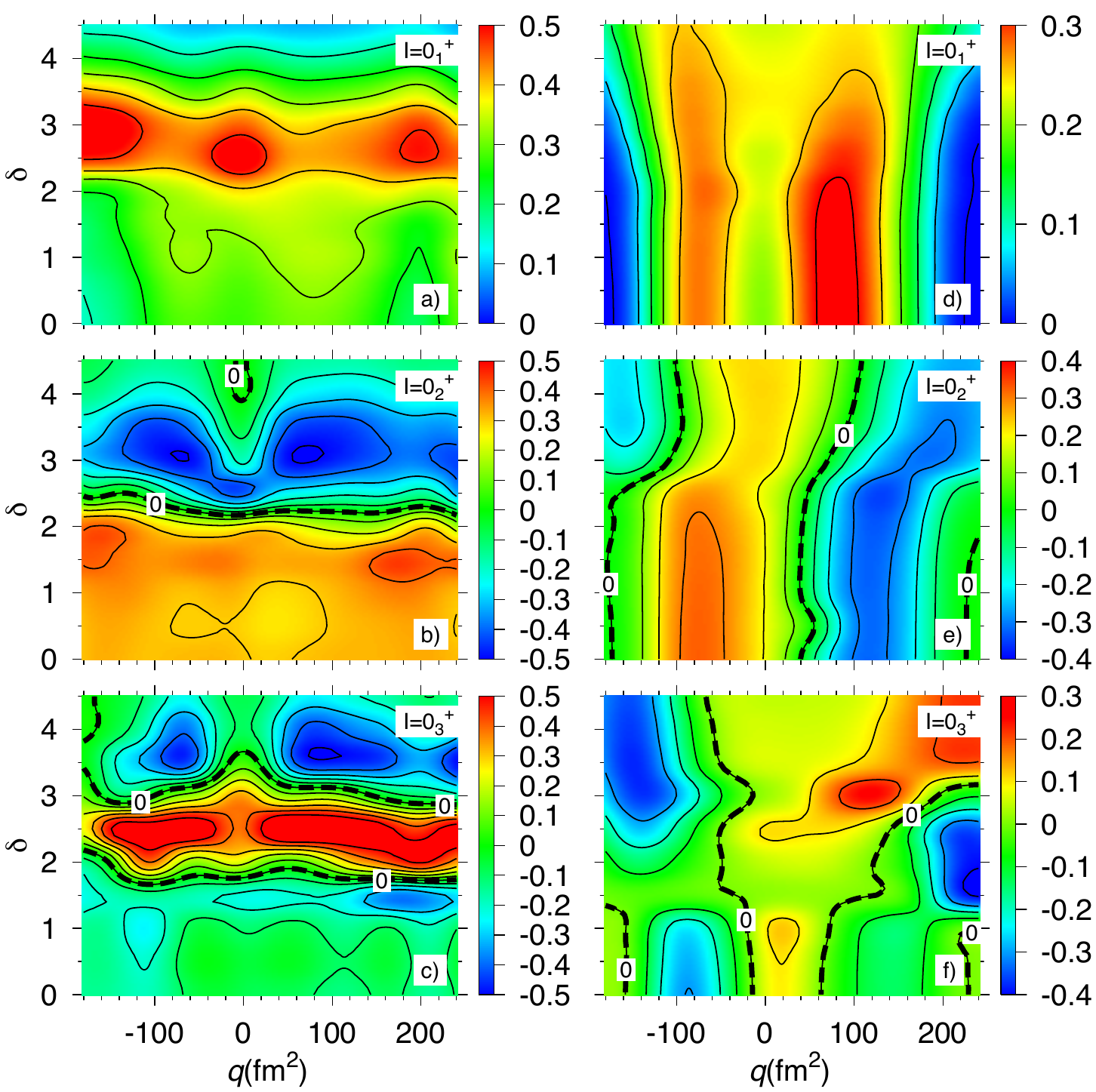}
\caption{ (Color Online) Wave functions of the lowest $0^+$ states of $^{52}$Ti of combined 1D generator coordinate calculations, see main text for further explanations. Left: Calculations for a fixed $q$-value taking  $\delta$ as generator coordinate.
Right: Calculations for a fixed $\delta$-value taking  $q$ as generator coordinate}
\label{pai_vib_2}
\end{center}
\end{figure}

In order to disentangle the role of the deformations and pairing vibrations we have performed two 1D calculations to decouple these degrees of freedom.
In the first one we do not consider the interaction of different nuclear shapes, i.e.,   we assume 
\begin{equation}
\langle q_1 \delta_1 | H \hat{P}^N\hat{P}^I|  q_2 \delta_2 \rangle  = \langle q_1 \delta_1 | H \hat{P}^N\hat{P}^I |  q_1 \delta_2 \rangle \delta_{q_2q_1}.
\end{equation}
For a Schr\"odinger type equation this would amount to diagonalize separately each $q$ value. For a HW type equation, however, additionally one has to assume 
\begin{equation}
\langle q_1 \delta_1  | \hat{P}^N\hat{P}^I|  q_2 \delta_2 \rangle  = \langle q_1 \delta_1  | \hat{P}^N\hat{P}^I|q_1 \delta_2 \rangle \delta_{q_2q_1}.
\end{equation}
to have that equivalence.
In any case, what we do is to perform separately, for each of the 32  $q$ values  of our 2D GCM basis, a 1D GCM calculation with $\delta$ as generator.  The two top plots of   Fig.~\ref{pai_vib_1}
correspond to two of such calculations. Now we take the lowest state of each of the the 32 values, i.e., the  $0^+_1$ state, and merge their wave functions in a 2D plot. The result of  drawing contour plots  is displayed in  panel (a) of Fig.~\ref{pai_vib_2}. The same is done with the  $0^+_2$  and  $0^+_3$ wave functions and plotted in the  panels (b) and (c)  of Fig.~\ref{pai_vib_2}, respectivelly.   The horizontal node line  in the $0^+_2$ plot, panel (b) indicates  that we have to do with a genuine pairing vibration. The plot on panel (a) with no node line corresponds to the ground state and the one in
(c) to a two phonon pairing vibration.

We now perform the other 1D calculation. In this case we do not allow  to interact states with different pairing energies,
that means
\begin{equation}
\langle q_1 \delta_1 | H  \hat{P}^N\hat{P}^I|  q_2 \delta_2 \rangle  = \langle q_1 \delta_1 | H  \hat{P}^N\hat{P}^I|  q_2 \delta_1 \rangle \delta_{\delta_2\delta_1}.
\end{equation}
and we have to do with 1D calculations, for the 10 fixed $\delta$ values of our basis, and the quadrupole degree as generator coordinate. In the lower panels of Fig.~\ref{pai_vib_1} we show two of such calculations, namely for $\delta=1.5$ and  $\delta=3.5$. We  observe that neither the potentials nor the  wave functions are harmonic.
As before we merge the 10 wave functions in a 2D plot for the $0^+_1$, $0^+_2$ and $0^+_3$ states as depicted in the  panels (d), (e) and (f) of  Fig.~\ref{pai_vib_2}, respectively.   Again the node lines are indicative of zero, one and two vibrational $\beta$  phonons.

The goal is to learn from the comparison of the {\em fake} 2D w.f. of Fig.~\ref{pai_vib_2} and the {\em real} ones
of  Fig.~\ref{fig:WF_J0_52Ti}, but before a consideration must be done. In the  plots of the {\em fake} 2D
wave functions  the normalization of the wave functions is different than in the {\em real} 2D calculations. For fixed
$q$, $\sum_{\delta} |g(q\delta)|^2 =1$, and for fixed $\delta$, $\sum_{q} |g(q\delta)|^2 =1$, while in the
full 2D calculations $\sum_{\delta,q} |g(q\delta)|^2 =1$. That means the wave functions will be transversally ( longitudinally ) extended in the calculation at fixed $\delta$ ($q$). 
This explains, for instance, the presence of strength
at  very large  pairing correlations for the $\delta$ fixed calculations. A look at the lower panels of 
Fig.~\ref{pai_vib_1} shows that the potential energy of the $\delta =3.5$ value is a few MeV higher in energy
than the $\delta =1.5$, indicating that as soon as pairing correlations are switched on the w.f. will be 
inhibited for very large $\delta$ values. That means, in these plots the energy consideration which will favor (hinder)    some values of $q$ or $\delta$  is not taken into account. For instance, maxima of the wave functions with very large
values of $\delta$ are highly unlikely to prevail.

We now proceed to compare these plots with the wave functions of the full 2D calculations displayed in the left
panels of  Fig.~\ref{fig:WF_J0_52Ti}.  The ground state of the full 2D calculations, see panel (a), presents two maxima,  a big one at
$+80$ fm$^2$ a smaller one at $-80$ fm$^2$. The bulk structure of this constellation is found in the ground state of the calculations with $q$ as generator coordinate, see  panel  (d) of Fig.~\ref{pai_vib_2}. We notice  here that,
as mentioned, the maxima are more extended in the $\delta$ coordinate  than the former ones. 
The ground state wave function with the pairing correlations as generator coordinate,  see panel (a) of Fig.~\ref{pai_vib_2},  is however completely different. It displays two maxima centered at  very large deformations, $\pm 200$ fm$^2$ corresponding to the large  level density which favors large pairing correlations and one at $q = 0$.  This w.f.  has little in common with the {\em real} 2D wave function. However, looking at this wave function at small and large pairing correlations, we can easily imagine that the consideration of the pairing interaction will modulate the ground  state w.f. of the  fixed $\delta$   calculations of panel (d).  

The $0^+_2$ state of the full 2D calculations, see panel (b) of  Fig.~\ref{fig:WF_J0_52Ti}, from the node structure point of view unequivocally  resembles the  $0^+_2$ of the calculations with $q$ as generator coordinate, see panel (e) of  Fig.~\ref{pai_vib_2}, except for the artificial  extension for large $\delta$ mentioned above.

The genuine pairing vibration of panel (b) of Fig.~\ref{pai_vib_2} does not have, however, a direct counterpart in the
second excited state of the full 2D calculations shown in panel (c) of 
Fig.~\ref{fig:WF_J0_52Ti}. On the other hand if we look at the self-consistent 1D calculations of Fig.~\ref{fig:WF_1D_Ti_signo}(a), we observe that the peak structure of the $0^+_3$ state (dottet line) is very similar to a cut along $\delta= 2.5$  in panel (f) of Fig.~\ref{pai_vib_2}. The value $\delta= 2.5$ is not arbitrary since corresponds approximately to the self-consistent path in panel (f) of Fig.~\ref{fig:pairfluc_Ti52}.
The state $0^{+}_{3}$ of Fig.~\ref{pai_vib_2}(f)  corresponds to a
two phonon $\beta$ vibration. Our claim, as we will justify below, is that the $0^+_3$ state of the full 2D calculations of 
Fig.~\ref{fig:WF_J0_52Ti} is a mixture of a  two phonon $\beta$ vibration and a pairing vibration.
The bulk structure of the two phonon $\beta$ vibration,
 see panel (f) Fig.~\ref{pai_vib_2}, have four peaks at approximate $q$ values $-140, -90, +100$ and $+220$ fm$^2$.
 The bulk structure of the one phonon pairing vibration of panel (b) of  Fig.~\ref{pai_vib_2}, consists in two peaks at 
$ -160$ and $+180$ and other two at $-50$ and $+50$ fm$^2$.  It seems that the most energetically efficient way to combine the node structure of the  two $\beta$ phonons and of the one pairing phonon is to favor the two peaks at large deformations (to profit from the large density level) with large pairing correlations  and the two at smaller deformation and smaller pairing correlations. That means the two oblate (prolate) peaks at large deformations stemming from both modes merge together and the same happens for the peaks at smaller deformations. The result is the $0^+_3$ state
of the full 2D calculation.

  From the above discussion we conclude that the quadrupole degree of freedom provides the bulk structure of
  the wave functions. The locations of the minima in the $q$-coordinate being modestly influenced by the pairing degree of freedom.  If we look at the 1D self-consistent path along the $(\delta, \beta)$ plane in 
   Fig.~\ref{fig:pairfluc_Ti52}(c) we observe little variation of the pairing content of the wave functions along this path, as a matter of fact it has a almost constant value of $\delta\approx  2.5$. This is not the case for the wave functions of the full 2D calculations, in particular for  the $0^+_3$ state,  the candidate  to the pairing vibration.
   What we infer from  Fig.~\ref{fig:WF_J0_52Ti} and  Fig.~\ref{pai_vib_2} is that the one phonon pairing vibration mix with the two phonon $\beta$ vibration as to accommodate pieces of the resulting wave function to different pairing strength.
   
 The conclusion of this part is that the presence of genuine pairing vibration is strongly hindered by the present of minima in the quadrupole degree of freedom which severely modify the node structure of the wave function. The presence of genuine pairing vibration should be limited to double shell closed nuclei and its closest neighborhood.

\section{Particle number distribution in the HFB+AMP approach.}

\label{part_dist}
An special aspect of the HFB+AMP approach is the particle number conservation. In this theory the particle number is adjusted, on the average, in the HFB wave functions by means of
constraints through Lagrange parameters. At the GCM level a correction is performed by 
the term introduced in Eq.~\ref{lamb_correct}.  The question we are interested in is: how good 
is the particle number conservation for the states $|\Psi^{I,\sigma}\rangle$? A direct answer to this question is provided by the particle number distribution of the wave function $|\Psi^{I,\sigma}\rangle$ . The probability
to find an eigenstate of $\hat{N}$ with eigenvalue $N$  in $|\Psi^{I,\sigma}\rangle$ is   given  by

\begin{eqnarray}
W^{N} &=& |\langle N |\Psi^{I,\sigma}\rangle|^2 = \langle\Psi^{I,\sigma} | N \rangle\langle N | \Psi^{I,\sigma} \rangle
\nonumber \\
&=& \langle \Psi^{I,\sigma} |P^{N} | \Psi^{I,\sigma} \rangle
\end{eqnarray}
In the same way $W^{Z,N} $ represents the probability to have simultaneously an eigenstate of $\hat{Z}$ and 
$\hat{N}$, with eigenvalues $Z$ and $N$ respectively. Taking into account the definition of $|\Psi^{I,\sigma}\rangle$, see Eq.~\ref{GCMstate}, one obtains
\begin{eqnarray}
 &&W^{Z,N}  =\langle \Psi^{I,\sigma}|P^{N}P^{Z}|\Psi^{I,\sigma}\rangle = \int dq dq^{\prime} d\delta d\delta^{\prime} \nonumber \\
&  &   \times f^{* b,\sigma}(q,\delta)
        \langle \phi(q,\delta)|\hat{P}^{I}\hat{P}^{N}\hat{P}^{Z}| \phi(q^{\prime} ,\delta^{\prime} )   f^{b,\sigma}(q^{\prime},\delta^{\prime})
\label{pesos}
\end{eqnarray}
in terms of known quantities.
Since the wave function $|\Psi^{I,\sigma}\rangle$ is normalized to the unity and $\sum_{N}P^{N}=\sum_{N}|N\rangle\langle N|=1$ is obvious that $\sum_{N,Z}W^{Z,N}=1$.

\begin{figure}[bht]
\begin{center}
\includegraphics[angle=0, scale=0.35]{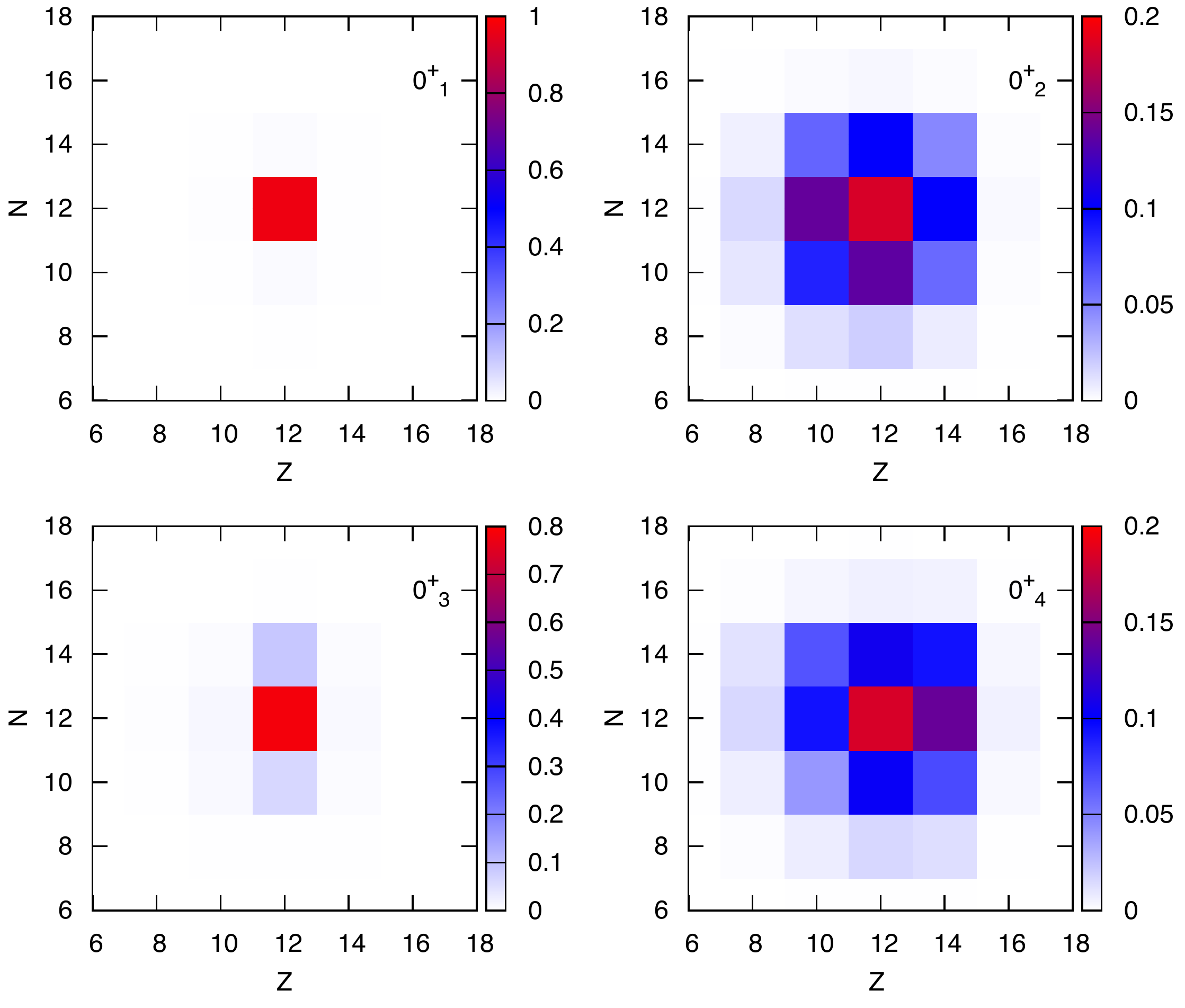}
\caption{(Color Online) Particle number distribution in the  ground and lowest excited states for $^{24}$Mg  states in 1D in the HFB+AMP approach.}
\label{fig:distri_N_1D}
\end{center}
\end{figure}
\begin{figure}[h]
\begin{center}
\includegraphics[angle=0, scale=0.35]{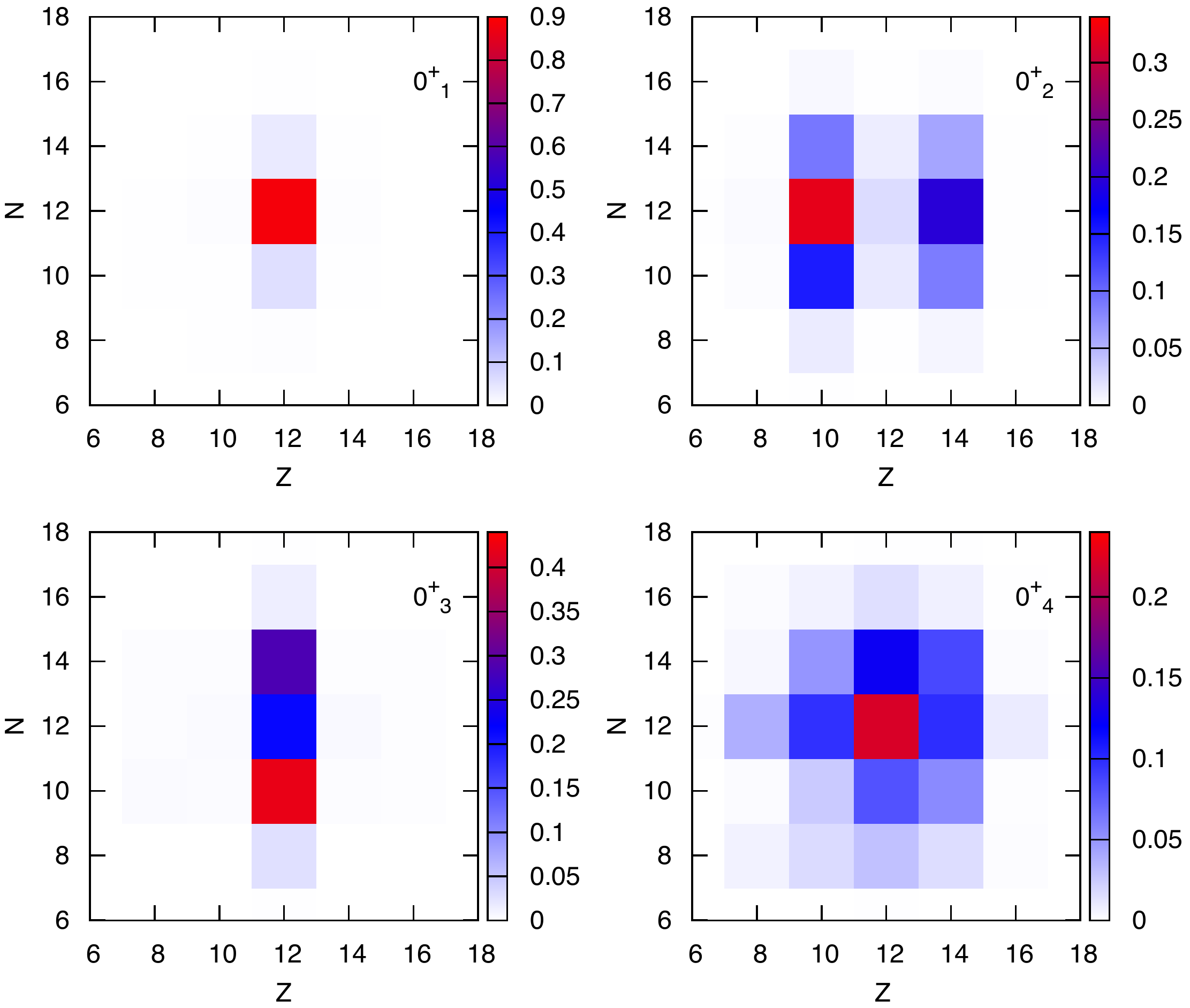}
\caption{(Color Online) Particle number distribution in the  ground and lowest excited states with $I=0^+\;\hbar$ for $^{24}$Mg  states in 2D for the HFB+AMP approach.}
\label{fig:distri_N_2D}
\end{center}
\end{figure}

  The usual picture that one has in mind \cite{ring} for such  distribution is that of a  BCS wave function and only for protons or neutrons. In this case one obtains a Gaussian distribution centered around the value $N$, where $N$ corresponds to the
constraint $\langle BCS|\hat{N}|BCS\rangle=N$ imposed in the solution of the variational BCS equations. 
In Fig.  ~\ref{fig:distri_N_1D} we show as an example the distribution $W(Z,N)$  for the  nucleus $^{24}$Mg and in the 1D case, that means with only $q$ as coordinate, for $I=0^+$ and for the four  lowest eigenstates of the HW equation. In the X -axis (Y-axis) we represent the number of protons (neutrons).  The color code on the right hand side of each plot indicates the value of $W(Z,N)$ for each set $(Z,N)$.  In this case the ideal distribution would be a 2D Gaussian distribution centered around $Z=N=12$. 
 For the ground state $0^+_1$ we find an almost pure eigenstate of $\hat{Z}$ and  $\hat{N}$. A  look at the 
 potential well of this nucleus, see panel (c) of Fig.~\ref{fig:1D}, indicates that the 
 wave function of this state will peak at around $q \approx 60$ fm$^2$, which corresponds (see  panel (d) of 
Fig.~\ref{fig:1D}) to zero pairing correlations. The distribution of the first excited state $0^+_2$ is shown in the top right panel, here we find a  very asymmetric distribution with respect to the line $N+Z=24$. The distribution of the state
$0^+_3$ indicates again that we are mainly confronted with zero pairing whereas the $0^+_4$ distribution is again far from Gaussian. In Fig.~\ref{fig:distri_N_2D} we present the same quantity for the two coordinates $(q,\delta)$ case and for the same
states. The distribution for the $0^+_1$  state is similar to the 1D case, for the  $0^+_2$ we obtain a distribution with a hole at $Z=N=12$ and a very asymmetric distribution. For the $0^+_3$ the maximum is at the wrong number of neutrons and 
for the  $0^+_4$ state though centered around the right value it is not a good Gaussian either. In most of the nuclei the results
are similar to these ones.

\section{Miscellaneous calculations in the PN-VAP+PNAMP approach}
\label{misc}
In this Section we apply the   theory described above to  calculate several relevant observables and study the impact of the pairing fluctuations.  Since we are mainly interested only in the relevance of the pairing fluctuations we will not discuss thoroughly 
the physics of the different issues. In the calculations we present results only in the PN-VAP+PNAMP approach.

\subsection{Separation energies}
\begin{figure}[h]
\begin{center}
\includegraphics[angle=0, scale=0.6]{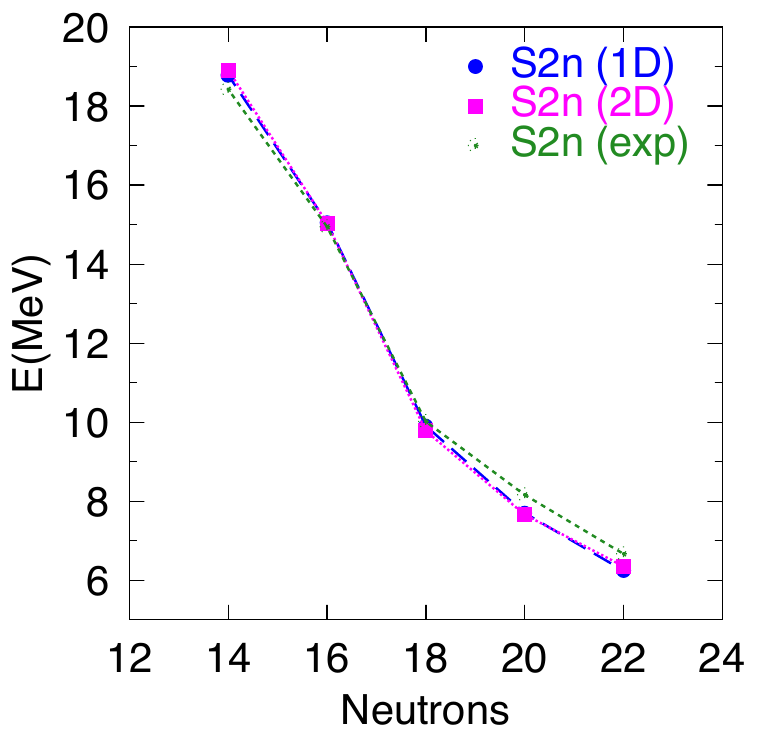} 
\caption{(Color Online) Two neutron separation energies ($S_{2n}$) in MeV for the Magnesium isotopes. The  experimental data are taken from Ref.~\cite{exp}}
\label{fig:S2n}
\end{center}
\end{figure}

An interesting quantity is the separation energy. This observable is the difference of two ground states energies, we do not expect therefore big changes  by the explicit consideration of the pairing degree of freedom.   In Fig.~\ref{fig:S2n} we
display the two neutron separation energies  for some magnesium isotopes in the 1D and 2D calculations as  compared with the experimental data. They are defined as the difference between the binding energies:  
\begin{equation}
S_{2n}(N)=B(Z,N)-B(Z,N-2)
\end{equation}

As expected the differences between the 1D and 2D are small and both of them show good agreement wit the 
experimental data.

\subsection {Electric monopole transitions $(E0)$} 
The E0 operator is given by
\begin{equation}
  \hat{T}({\rm E0}) =  \sum_k e_k \hat{r}_{k}^2 
  \label{E0_oper}
\end{equation}
The diagonal matrix element provide the charge radius and the non-diagonal are related with transitions. Since
the radius is strongly related with the shape of the nucleus, the corresponding $E0$ transitions are also related
with the shape of the initial and final states, see Ref.~\cite{Wood1999323,RevModPhys.83.1467}.  For example the $E0(0^+_2 \longrightarrow 0^+_1)$ allows to differentiate limiting situations in which two configurations compete for the ground and first excited state. Thus in the island of inversion the deformed configuration based on two neutrons being excited from the d$_{3/2}$ to the intruder orbital f$_{7/2}$ keeps pace with the normal spherical one as illustrated by the well-known case of $^{32}$Mg where the intruder state even becomes the ground state, 
Ref.~\cite{PhysRevLett.103.012501}. In such a situation of competing configurations and in the absence of mixing one expects either a deformed $0^+_1$ and a nearly spherical 
$0^+_2$ state or the other way around. The transition probability is given by
\begin{equation}
   \rho^2({E0}) = \frac{1}{R^4} \left| \langle \Psi_f | \sum_k e_k r_{k}^2 | \Psi_i \rangle \right|^2.
\label{maelem}
\end{equation}
with $R = 1.2$A$^{1/3}$,  $\Psi_i$ and $\Psi_f$ are the wave functions of the initial and final nuclear states, in our case the 
$0^+_2$ and $0^+_1$ states. As we have seen in the earlier sections the $0^+$ states are influenced by pairing correlations, thus we expect some differences between the 1D and 2D calculations. 

\begin{figure}[h]
\begin{center}
\includegraphics[angle=0, scale=0.55]{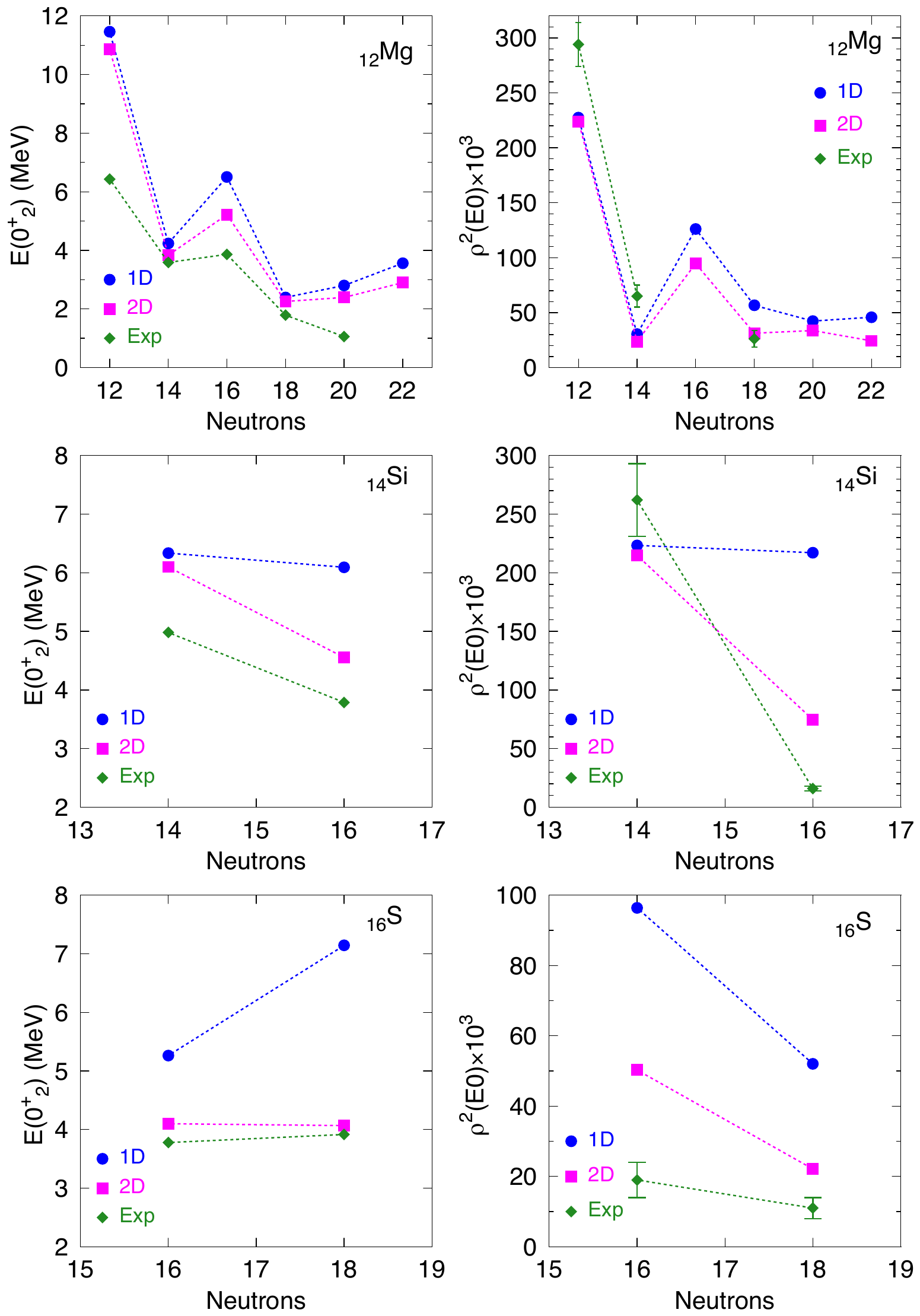} 
\caption{ (Color Online) Excitation energies for the $0^{+}_{2}$ states and $E0$ strength $\rho^2($E0$,0^+_2 \rightarrow 0^+_1)$
for the Mg isotopes (top panels), the Si isotopes (medium panels) and S isotopes (bottom panels).  The experimental values are taken from 
\cite{Kibedi200577,exp,Wood1999323,PhysRevLett.103.012501}. }
\label{fig:E0_all}
\end{center}
\end{figure}
  
In Fig.~\ref{fig:E0_all}, top panels,  we display the  excitation energies and the E$0(0^+_2 \longrightarrow 0^+_1)$ values for the
Magnesium isotopes where experimental information is available. Qualitatively there is not much difference between the 1D and 2D predictions for the excitation energy of the $0^+_2$ state, the maximum difference being circa 1 MeV. The 
  experimental tendency is fairly well reproduced in both approaches though the 2D improves somewhat the agreement with the experiment. The discrepant   $^{24}$Mg
value is  probably due to the fact that this state is a rather pure two quasiparticle state. In the right top panel the $\rho^2(E0, 0^+_2\longrightarrow0^+_1)$ values for the same nuclei are shown. Again there is not a qualitative difference between both predictions. Both display very well the experimental behavior though the 2D calculations predict lower values than the 1D one.  In the particular case of $^{30}$Mg the  2D calculations reduce the 1D value by a factor of two in such a way that the experimental value is correctly reproduced.

 In Table \ref{table2} the charge radius is presented for the Magnesium isotopes, in the two approaches and compared with the experiment. Both calculation  differ slightly and both show good agreement with the experimental values.

In the medium panels of Fig.~\ref{fig:E0_all} we present the excitation energy of the $0^+_2$ states and the monopole transition probability for the Si isotopes.
 In these nuclei we do observe qualitative and quantitative differences between the 1D and the 2D calculations. 
Whereas the 1D predictions do not provide neither the tendency nor the right value,  the 2D calculations improve considerably the agreement  with the experimental values.  For  $^{30}$Si the improvement is spectacular, the excitation energy of the $0^+_2$ state is reduced in approximately 1.5 MeV to reproduce the experimental tendency.  With respect to the monopole strength, the 2D is about a factor three  smaller than the 1D, getting closer to the experimental value. These large changes have probably to do with the 2s$_{1/2}$ sub-shell closure for $N=16$.  

Finally, in the bottom  panels of Fig.~\ref{fig:E0_all} we display the results for the Sulfur isotopes. Concerning the energy of the 
$0^+_2$ states we find that as with the Si isotopes, the 1D predictions describe very poorly the data. The consideration
of the pairing fluctuations again reduce considerably these values as to reproduce very good the experimental values. The same
can be said for the monopole strength, the 2D predictions reduce by a factor of two the 1D calculations in such a way that the data
are better reproduced. 

\begin{table}[h]
\begin{center}
\begin{tabular}{|c|c|c|c|c|c|}\hline
$\langle r^{2}\rangle ^{1/2}_ {\rm ch}$ ${\rm fm}^{2}$ & 1D  &  2D  & Exp \\ \hline \hline
${^{24}}$Mg &  3.095   &  3.098   &   3.057       \\\hline
${^{26}}$Mg &  3.065   &  3.068   &   3.034       \\\hline
${^{28}}$Mg &  3.078   &  3.082   &   3.070       \\\hline
${^{30}}$Mg &  3.106   &  3.110   &   3.111       \\\hline
${^{32}}$Mg &  3.158   &  3.159   &   3.186       \\\hline
${^{34}}$Mg &  3.210   &  3.213   &    \\\hline \hline
\end{tabular}
\end{center}
\caption{Nuclear charge radii  for the ground state of the Magnesium isotopes, the experimental values
are taken from  \cite{PhysRevLett.108.042504}}
\label{table2}
\end{table}

\subsection {Quadrupole $ E2( 0^+_1 \longrightarrow 2^+_1)$ transition probabilities}

\begin{figure}[t!]
\begin{center}
\includegraphics[angle=0, scale=0.55]{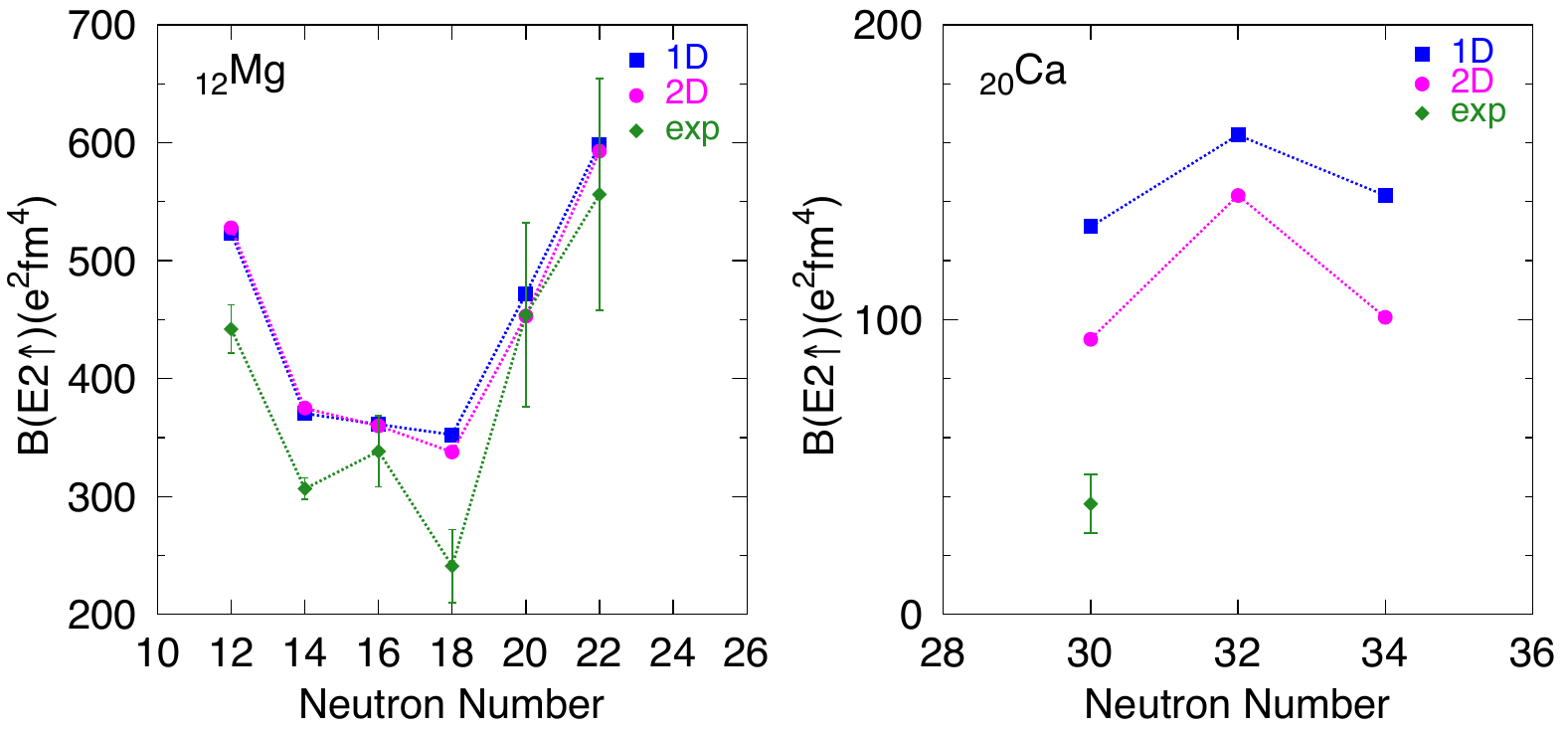} 
\caption{(Color Online) E2 transition probabilities for the Mg and Ca. Experimental data are taken from \cite{exp,PhysRevC.71.041302,PhysRevLett.102.242502,PhysRevLett.103.012501}}
\label{fig:BE2}
\end{center}
\end{figure}

 In Fig.~\ref{fig:BE2} the reduced transition probabilities $B(E2, 0^+_1 \longrightarrow 2^+_1)$,  for some of the nuclei earlier discussed,  are displayed.  In the left 
 panel for the Magnesium and in the right one for the Calcium isotopes. The theoretical predictions in the 1D approach for the Mg chain  reproduce qualitatively well the experimental behavior despite the fact they are a somewhat  larger.  The inclusion of the pairing fluctuations, in general, makes the 2D predictions a bit smaller and therefore closer to the experimental data. For the calcium isotopes we find larger contributions of the pairing fluctuations, they amount to a reduction of $30 \%$ of its 1D values.  In the case of the nucleus $^{52}$Ti in the 1D calculations one obtains 
a  $B(E2, 0^+_1 \longrightarrow 2^+_1)$ of  643.3 $e^{2}$ fm$^{4}$, in the 2D 601.2  $e^{2}$ fm$^{4}$ to be compared
with the experimental value of 567 (51) $e^{2}$ fm$^{4}$.  The behavior goes in the same lines as before: the pairing fluctuations reduce the, otherwise, to large values bringing the predictions closer to the experimental values.

\section{Conclusions}

   In this work we have performed a thorough research of the pairing degree of freedom in beyond mean field theories. 
  The quality of the interaction namely the finite range density dependent Gogny forces guarantees  the proper
  treatment of the pairing correlations.
  
   We have shown that the consideration at the same footing of the pairing degree of freedom and the quadrupole
deformation at the different stages of the calculations provides a considerable improvement of the description of
many observables of atomic nuclei. 
   
   We have underlined the importance of the conservation of symmetries, in particular we have shown the superiority 
   of the variation after projection for the particle number case as compared with the  plain HFB approach and the projection after variation one.  This supremacy manifests itself at the different levels, from the simplest to the
  most sophisticated ones in particular in the solution of the Hill-Wheeler equation with the quadrupole and the pairing degrees of freedom as generator coordinates. We have found in particular that the absence of particle number projection leads to a larger linear dependence implying thereby smaller variational spaces and an unnatural 
  strength concentration. As a consequence the spectra became more compressed that the particle number projected
  counterparts.
  
  The role of the pairing degree of freedom has been analyzed in the most sophisticated approach (VAP-PN+PNAMP)
with a large number of observables: spectra, $E0$ and $E2$ transition probabilities, separation energies among others. In all studied nuclei we find a better agreement with the experimental data  as compared to simpler theories. The pairing
vibrations are also thoroughly discussed with the finding that the quadrupole degree of freedom strongly inhibits the
presence of genuine pairing vibration.

The validity of our calculations is limited  by the absence of explicit single particle degrees of freedom (two-quasiparticle excitations) as well as the restriction to axial symmetry. The small size of our configuration space also limit the accuracy of our prediction.

In the present work we have analyzed light  nuclei mainly in the vicinity of shells closures in order to keep the
 configuration space small. The analysis of  heavier,  spherical and  deformed nuclei  will be done for some particular cases in a following paper. We also plan to perform separate constraints on the proton and neutron pairing gap to study specific aspects which are genuine for protons or neutrons.

\section{Acknowledgments}
The authors acknowledge financial support from the Spanish Ministerio de  Ciencia e Innovaci\'on under contracts FPA2011-29854-C04-04, by the Spanish Consolider-Ingenio 2010 Programme CPAN (CSD2007-00042). N.L.V acknowledges a scholarship of the Programa de Formaci\'on de Personal  Investigador (Ref. BES-2010-033107). T.R.R. acknowledges support from BMBF-Verbundforschungsprojekt number 06DA7047I and Helmholtz International Center for FAIR program.

\section{Appendix: Details of the density dependent term.} 

In Ref.\cite{Anguiano2001a} it  was shown that in calculations with  a density dependent interaction and in a
particle number projected  approach there are two sources for divergencies.  The first one
is connected  with the neglecting  of exchange terms. Obviously, these divergencies can  be straighten out
by including {\em all} missing exchange terms of the interaction. 
The second one has its origin in the density dependent term of the interaction which we shall call $V_{DD}$. This term was conceived for  plain mean field approaches where only expectation values, i.e., diagonal matrix elements, do appear. Consequently in the mean field approach $V_{DD}$ is constructed to depend on the mean field density.
  In theories beyond mean field, for example in particle number projection, the contribution to the energy of the 
 density dependent term is given by
\begin{equation}
E^P_{DD}    =  \frac{\langle \Phi^N | {\hat{V}}_{DD}\left [ \overline{\rho} 
(\vec{r}) \right ] | \Phi^N \rangle} 
{\langle \Phi^N | \Phi^{N} \rangle} \nonumber \\
 =  \frac{  {\displaystyle 
\int }   d {\varphi}_{} \langle 
\phi  | {\hat{V}}_{DD}
 \left [ \overline{\rho} (\vec{r}) \right ]  
e^{i{\varphi}_{} {\hat{N}}_{} } | \phi \rangle  } 
{ {\displaystyle  \int } 
 d {\varphi}_{} \langle 
\phi | 
e^{i{\varphi}_{} {\hat{N}}_{} } | \phi \rangle }
\label{eq:epdd}
\end{equation}
where $\left [ \overline{\rho} (\vec{r}) \right ]$ indicates the explicit
dependence of $V_{DD}$ on a density $ \overline{\rho} (\vec{r})$ to be specified. Looking
at these expressions it is not obvious which dependence should be used.
There are two more or less straightforward prescriptions \cite{Valor200046}  for 
$ \overline{\rho} (\vec{r})$~:

\begin{figure}[t]
\begin{center}
\includegraphics[angle=0, scale=0.6]{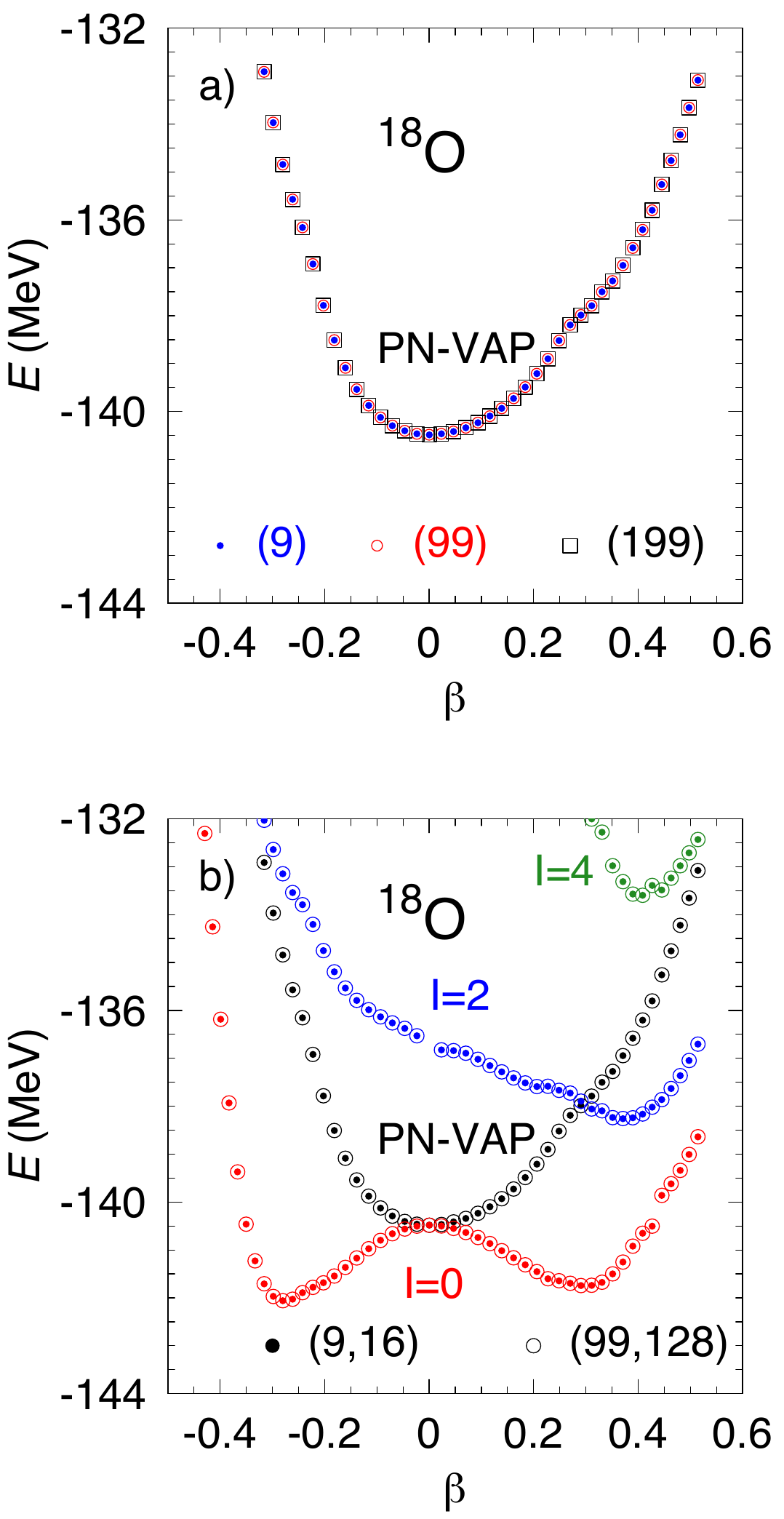}
\caption{ a) Potential energy surfaces in the PN-VAP approach for different number of integration points.
b) Potential energy surfaces in the PN-VAP approach and  in the PNAMP approach for $I=0, 2$ and 4 for different number of integration points. }
\label{poles}
\end{center}
\end{figure}

The first prescription is inspired by the following consideration~:
In the mean field approximation, the energy is given by
$\langle \phi | \hat{H} | \phi \rangle /{\langle \phi | \phi \rangle}$
and $V_{DD}$ is assumed to  depend on the density  
$\langle \phi | \hat{\rho} | \phi \rangle/{\langle \phi | \phi \rangle}$.
On the other hand, if the wave function which describes the nuclear system 
is the projected wave function $| \Phi^N \rangle$,   we 
have to calculate the matrix element
$\langle \Phi^N | \hat{V}_{DD} | \Phi^N \rangle /{\langle \Phi^N | \Phi^N \rangle}$
(see the middle term in eq.~(\ref{eq:epdd})).
It seems reasonable, therefore,  to use in $V_{DD}$ the density 
$ \overline{\rho} (\vec{r})\equiv \rho^N(\vec{r})= \langle \Phi^N | \hat{\rho} | \Phi^N \rangle /
{\langle \Phi^N | \Phi^{N} \rangle} $, i.e. the projected 
density. 
One has to be aware that this prescription  can only  be used in the case of the particle number 
projection where one projects in the gauge space associated to the particle number operator and which has nothing to do
with the spacial coordinates. In the case of symmetries associated with $\vec{r}$ like the angular momentum or parity projection one has to work with the second prescription.

 The second prescription has been guided by the choice usually done in the
Generator Coordinate method with density dependent forces \cite{Bonche1990466}. The
philosophy behind this prescription is the following: to
evaluate eq.~(\ref{eq:epdd}) we have to calculate matrix elements between different
product wave functions $|\phi \rangle$ and $| \tilde{\phi} \rangle$
($|\tilde{\phi} \rangle =e^{i\varphi\hat{N}}| \phi \rangle$)
(see last term in eq.~(\ref{eq:epdd})). 
Then, to calculate  matrix elements of the form
${\langle \phi | {\hat{V}}_{DD} | \tilde{\phi} \rangle}{\langle \phi |\tilde{\phi} \rangle}$
we choose the mixed density
  $\overline{\rho} (\vec{r})= {\rho}_{\varphi} (\vec{r}) =
 {\langle \phi | \hat{\rho} (\vec{r})| \tilde{\phi} \rangle}{\langle \phi |
\tilde{\phi} \rangle}$
to be used in $ {\hat{V}}_{DD}$.
This approach is called the mixed density prescription. \\

Both prescriptions have been tested with the Gogny force in the Lipkin
Nogami approach \cite{Valor1997249} and practically no difference was found in the 
numerical applications. One should notice that in the second prescription 
$\overline{\rho} (\vec{r})$ depends on the angle $\varphi$ at variance with the first
prescription.

It has been shown in ref.~\cite{Anguiano2001a} for the particle number projection that  the projected prescription is free from
divergences while the mixed prescription may present some problems.

In order to get rid of the divergencies in our calculations we include {\em all} exchange terms of the Gogny force. 
Concerning the density dependent term we use the projected density in the particle number projection case and the mixed density prescription in the angular momentum projection and in the GCM cases.

To illustrate the absence of divergences under these conditions we discuss  the paradigmatic case of $^{18}$O used by Bender and collaborators \cite{PhysRevC.79.044319} to show the presence of divergencies in the case of the Skyrme force and without taking care to remedy the above mentioned problems. In Fig.~1 of Ref.~\cite{PhysRevC.79.044319} the potential energy of the nucleus $^{18}$O was plotted agains the $\beta$ degree of freedom taking 5 and 199 integration points in the discretization of the projection of the number of particles. In this figure one can observe two poles at  $\beta \approx 0.22$ and $\beta \approx -0.3$ for 199 points.   We now present in Fig.~\ref{poles} (a) the same plot for the Gogny force calculated in the way mentioned above taking 9, 99 and 199 integration points. One can immediately observe not only the absence of any divergency but also the perfect convergence of the calculations.

Concerning the angular momentum projection the current status is that the studies on particle number projection  \cite{PhysRevC.79.044319,PhysRevC.76.054315} have been simply extrapolated to this case. However, to our knowledge, an explicit study of the existence of divergences and/or steps by the use of the mixed density prescription in the AMP case has not been carried out.   Even more, we have looked explicitly for such an ill-behavior in many calculations with axial and triaxial angular momentum projection and we have never found any hint of them. As an example, in Fig.~\ref{poles} (b) we also show the results for simultaneous projection of particle number and angular momentum for  $I = 0, 2$ and $4\; \hbar$.  We present
two calculations, one with 9 (16) integration points for  particle number (angular momentum) and another
with 99 (128), respectively.  Again, in contrast to Refs~\cite{PhysRevC.79.044319,PhysRevC.76.054315}, the absence of divergences/steps and the good convergence are manifest.

The conclusion  with respect to the use of the mixed density prescription in the case of the angular momentum projection with the Gogny interaction is that either there are no problems with its non-analicity in the complex plane or they appear, contrary to the PNP case, so seldom that the probability of finding them in practical calculations is quite negligible.  Two last concluding remarks: First, obviously a detailed studied  should be performed in order to clarify this issue, and second, our conclusion is limited to the present type of calculations, i.e.,  without time reversal breaking.

\bibliography{artmini}
										
\end{document}